\documentclass{aastex}
\usepackage{natbib}
\usepackage{icarusbib}
\usepackage{lastpage}

\newcommand{\simlt}{\lower.5ex\hbox{$\; \buildrel < \over \sim \;$}}

\shorttitle{Debris and Fragments of Comet 73P}
\shortauthors{Reach {\it et al.}}

\begin{document}

\title{Distribution and properties of fragments and debris from 
the split comet 73P/Schwassmann-Wachmann 3 as
revealed by {\it Spitzer} Space Telescope}

\author{William T. Reach}
\affil{Infrared Processing and Analysis Center, 
MS 220-6, California Institute of Technology,
Pasadena, CA 91125}
\email{reach@ipac.caltech.edu}

\author{Jeremie Vaubaillon}
\affil{Infrared Processing and Analysis Center, 
MS 220-6, California Institute of Technology,
Pasadena, CA 91125}

\author{Michael S. Kelley}
\affil{Department of Astronomy, University of Minnesota, Minneapolis, MN 55455}

\author{Carey M. Lisse}
\affil{Johns Hopkins Applied Physics Laboratory, SD/SRE-MP3/E167, 11100 Johns Hopkins Road, Laurel, MD 20723}

\author{Mark V. Sykes}
\affil{Planetary Science Institute, 1700 E Ft. Lowell, Suite 106, Planetary Science Institute, Tucson, AZ 85719}

\slugcomment{Manuscript Pages: \pageref{LastPage},
Tables: \ref{sumtable}, 
Figures: \ref{profjan}
}

\noindent {\bf Proposed running head:} Debris and Fragments of Comet 73P
\bigskip

\noindent {\bf Editorial correspondence to:} 

Dr. William T. Reach

Infrared Processing and Analysis Center

MS 220-6

Caltech

Pasadena, CA 91125

Phone: 626-395-8565

Fax: 626-432-7484

E-mail: reach\@ipac.caltech.edu

\clearpage


During 2006 Mar - 2007 Jan,  we used the IRAC and MIPS instruments on the {\it Spitzer} Space Telescope to study the infrared emission from the ensemble of fragments, meteoroids, and dust tails in the more than 3 degree wide 73P/Schwassmann-Wachmann 3 debris field. We also investigated contemporaneous ground based and HST observations. 
In 2006 May, 55 fragments were detected in the {\it Spitzer} image.
The wide spread of fragments along the comet's orbit indicates they were formed from the 1995 splitting event.
While the number of major fragments in the {\it Spitzer} image is similar to that seen from the ground by optical
observers, the correspondence between the fragments with optical astrometry and those seen in the {\it Spitzer} images
cannot be readily established, due either to strong non-gravitational terms, astrometric uncertainties, or transience of the fragments
outgassing.
The {\it Spitzer} data resolve the structure of the dust comae at a resolution of $\sim 1000$ km, and they reveal the infrared emission due
to large (mm to cm size) particles in a continuous dust trail that closely follows the projected orbit. 
We detect fluorescence from outflowing CO$_{2}$ gas from the largest fragments (B and C), and we measure the CO$_2$:H$_2$O proportion (1:10 and 1:20, respectively). 
We use three dimensionless parameters to explain dynamics of the solid particles:
the rocket parameter $\alpha$ is the reaction force from day-side sublimation divided by solar gravity,
the radiation pressure parameter $\beta$ is the force due to solar radiation pressure divided by solar gravity, and
the ejection velocity parameter $\nu$ is the particle ejection speed divided by the orbital speed
of the comet at the time of ejection. The major fragments have $\nu>\alpha>\beta$ and are dominated by the
kinetic energy imparted to them by the fragmentation process. The small, ephemeral fragments seen by HST in the tails of the
major fragments have $\alpha>\nu>\beta$ and are dominated by rocket forces (until they become devolatilized). The meteoroids 
along the projected orbit seen by {\it Spitzer} have $\beta\sim\nu\gg\alpha$ and are dominated by radiation pressure and ejection velocity, though both influences are much less than gravity.
Dust in the fragments' tails has $\beta\gg(\nu+\alpha)$ and is dominated by radiation pressure.

\clearpage

\section{Introduction}

Comet 73P/Schwassmann-Wachmann 3 split in 1995,
with major fragmentation beginning on 1995 Sep 7 \citep[][]{sekanina}. 
Phenomena associated with the split
included dramatically increased OH gas emission \citep{crovisierSW3},
an optical brightness increase by 4 magnitudes,
and the presence of at least two and likely 4 fragments.
Two of the fragments (73P-B and 73P-C)
from the 1995 event
were long-lived and
returned on both the 2001 and 2006 perihelion passages. 
(Based on the comet's 5.39 yr period, the next perihelion passage will be in 2011 Oct.)
73P-C is the brightest fragment and `leads' 73P-B in the orbit, 
suggesting it is
the principal and most massive remnant of the pre-1995 comet, by comparison to
other split comets where the main fragment generally leads the others
\citep{boehnhardt}.
In the 1990 apparition, prior to the split, 73P was observed
as a single, intact comet,
with an upper limit on fragments being 3.2 mag fainter than the
primary \citep{jewittsplitting}.
Yoshida's compilation of the 
lightcurve\footnote{http://www.aerith.net/comet/catalog/0073P}
shows a typical brightness profile for a Jupiter-family comet in 1990
(with peak heliocentric magnitude 11.5),
and a nearly identical brightness trend being followed in 1995 up until 
1995 Aug, when it brightened by 6 magnitudes.
However, \citet{lisse98} reported significantly enhanced mid-infrared emission from the comet detected by the 
Diffuse Infrared Background Experiment on 
the {\it Cosmic Background Explorer} in 1990, suggesting increased emission of heavy dust particles.
On the 2001 return, fragments B and C were observed and followed a typical 
activity curve, elevated by approximately 3.5 mag with respect to that 
observed in 1990. At least one new fragment, 73P-E, was found in the 2001 return but
it quickly faded from view (and did not return in 2006).

In 2006, as the fragments of 73P again approached perihelion, there was renewed
interest due to 73P's anticipated close
approach to the Earth ($\Delta=0.08$ AU) on 2006 May 13, just before perihelion 
($q=0.939$ AU) on 2006 Jun 7. 
That the comet's orbit nearly intersects the Earth's is
significant not only because it becomes bright for ground-based observations
but also because the comet has an associated meteoroid stream, the
$\tau$ Herculids \citep{jenniskens}, and the comet 
is also an excellent mission target, as it can be reached
relatively inexpensively by spacecraft. 
(Indeed 73P was considered as a
target for the ROSETTA and CONTOUR missions.)
 Amateur and professional
observations revealed an extensive set of fragments along the orbit of 73P,
with the brightest fragment still being 73P-C 
(though in late 2006 Apr, 73P-B was brighter)
and some small fragments found ahead of 73P-C. 
Yoshida's compilation of lightcurves shows a typical shape for 73P-C, though
it remained some 2.5 mag brighter than 73P was in 1990. 
The lightcurve of fragment B has at least one significant increase, becoming
some 2.5 mag brighter beginning around 2006 Mar 31 and returning to its pre-flare-up
trend by 2006 May 8.
Fragment G was first observed in 2006 Feb and it had a significant
flare-up by approximately 2 magnitudes, beginning 2006 Apr 5 then
fading from view in early May.
The flare-ups of fragments B and G were almost certainly associated with significant 
sub-fragmentation as was observed by ground-based 
telescopes \citep[e.g.][]{fuse} and 
space-based observatories \citep[including HST and {\it Chandra};][]{weaverSW3,wolk08}.

By studying the
properties of the fragments of split comets, we hope to determine the properties of fresh surfaces
of comet nuclei as well as the structure of the nucleus.
Comet nuclei are heated to depths of several cm during diurnal 
(and of order meters for orbit-averaged) heat pulses, 
and they lose of order 1 m of their surface 
 each perihelion passage. When comets split into large fragments, fresh interior surfaces are exposed, allowing observation of relatively primitive material from the epoch of the comet's formation. 
The major fragments B and C were observed spectroscopically and found to be very similar to each other, both being carbon-chain depleted \citep{hitomi}.
The similarity of their composition is in contrast to the fact that fragment B was actively producing icy sub-fragments during the 2006 apparition while fragment C appeared to exhibit only `normal' activity without sub-fragmentation in 2006.
Thus the chemical properties of the ice may not be very different at depth 
than they are for active regions in the surface, but the distribution of
ice within the nucleus may be highly nonuniform.
In addition to the chemistry of the freshly-outgassing surfaces, we can
learn about the structure of cometary nuclei from the way they split.
Fragmentation will proceed rapidly for subunits that are weakly bound, possibly
leading down to the size of cometesimal building blocks. 
Debris of a wide range of sizes was inferred from observations of C/1999 S4 (LINEAR),
with comparable mass in large ($\sim 0.3$ m) meteoroids and detectable 
($\sim 40$ m) fragments \citep{weaverlinear,bockleemorvan01}.

In this paper we present {\it Spitzer} Space Telescope
mid-infrared observations of the fragments
and debris from 73P.
The outstanding sensitivity of space-based mid-infrared observations for
extended, inner solar system objects makes these observations unique
in their ability to simultaneously image the fragments, meteoroids, and dust.
In a companion paper, we analyze the dynamics of the dust and meteoroids to infer the
size distribution, finding that there is an excess in the abundance of mm-sized
and larger particles relative to a power-law fitted to the smaller particles
\citep{vaubaillSW3}. This paper concentrates on the 
physical properties and dynamics of the fragments.

\section{Observations}

\begin{figure}
\plotone{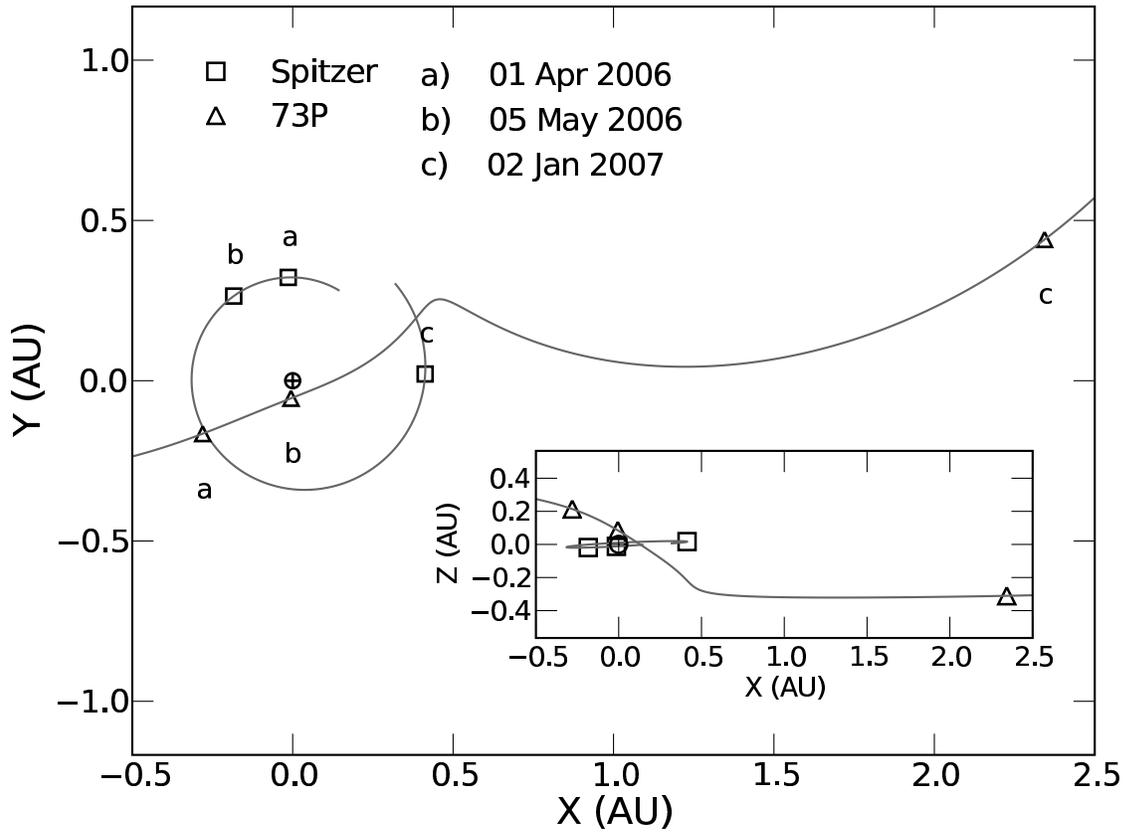}
\figcaption{
The relative positions of Spitzer ($\square$), Earth ($\oplus$),
and comet 73P ($\triangle$) on the dates of our MIPS observations.  The
positions are shown in a geocentric ecliptic coordinate system, and the
orbits of Spitzer and 73P are plotted as gray lines.
\label{obsfig}}
\end{figure}

More than 57 detected discrete nuclear fragments, associated meteoroids, and the anti-solar dust tails of 73P/SW-3 
span a complex debris field along more than 3 degrees of mean anomaly of the comet's orbit.
This 
debris field was mapped three
times as part of 3 {\it Spitzer} observing programs. All observations
were designed  to cover the region spanning from ahead of fragment
C to well behind fragment B.
Table~\ref{obslog} lists the observing conditions, including the length
of the arc along the orbit covered by each mosaic. To illustrate the viewing geometry at the 
times of the {\it Spitzer} observations, Figure~\ref{obsfig} shows the locations of Earth, {\it Spitzer},
and the comet.

\begin{deluxetable}{lrrccccccr}
\tablecaption{Log of observations\label{obslog}}
\tablewidth{7truein}
\tabletypesize{\footnotesize}
\tablehead{
\colhead{Start Date} & \colhead{$R$} & \colhead{$\Delta_{S}$~\tablenotemark{a} }
&  \colhead{Wavelengths} & \colhead{Arc\tablenotemark{b}} &
 \colhead{Tiles} & \colhead{$T_{tile}$\tablenotemark{c}} &  \colhead{$\delta T_{shad}$\tablenotemark{d}} & \colhead{$N_{image}$\tablenotemark{e}}\\
\colhead{(UT)} & (AU) & (AU) & ($\mu$m) &  ($10^{6}$ km) & ($N\times\arcmin\times \arcmin$) & (min) & (min) \\
}
\startdata
2006-03-28  09:35 & 1.37 & 0.64 & 3.6,4.5,5.8,8 & 3.4 & $3\times47\times 30$ & 40 & 122 & 1,854\\
2006-04-01  19:19 & 1.33 & 0.60 & 24, 70, 160    & 3.6 & $1\times124\times 82$ & 181& 325 & 5,200\\
2006-05-04 12:49 & 1.06  & 0.37  & 24, 70, 160   & 6.7 & $4\times131\times 84$ & 181 & 181 & 20,800\\
2007-01-01 17:16           & 2.80  & 2.58 &  24                  & 5.2 & $1\times46\times 8$ & 162 & ... & 672\\
\enddata
\tablenotetext{a}{Distance from the observatory ({\it Spitzer}) to the comet}
\tablenotetext{b}{Distance along the comet's orbit covered by the observation}
\tablenotetext{c}{Duration of one map tile}
\tablenotetext{d}{Interval between primary and shadow observations (for each tile)}
\tablenotetext{e}{Number of images ($5'\times 5'$ field of view for IRAC at each wavelength or MIPS at 24 $\mu$m)}
\end{deluxetable}

\subsection{2006 Mar-Apr}
The {\it Spitzer} Cycle 2 GO imaging observations (Program ID 20039) in 
2006 Mar-Apr were performed 
with the Infrared Array Camera \citep[IRAC][]{fazioirac} and 
the Multi-band Infrared Photometer for {\it Spitzer}
 \citep[MIPS][]{rieke04}, with very good viewing 
conditions---the comet was close to Earth and {\it Spitzer},
and the
line of sight was inclined by $20^{\circ}$ from the orbit plane 
(in contrast to common comet observing conditions 
at orbit plane angles $<5^{\circ}$). The map center and size were tuned
based on rapidly evolving knowledge of the fragment locations as of
2006 February, as gleaned from amateur astronomers' internet postings
and our own reconnaissance using the Palomar Observatory Hale telescope.
The images were made twice, with a separation of 5.4 hr between epochs;
during the time between epochs, the comet moved by 14.9$'$, so a completely
different background field of galaxies was present. 

We combined the images at the two epochs into a comet-only image as follows.
First, we registered the mosaics 
made for each epoch, so
that the comet fragments appear in the same pixels. 
Then, we created a celestial-background-subtracted image by taking, for each pixel,
the lowest of the two observed brightnesses (epochs 1 and 2).
This method works because the sky brightness is the sum of comet emission plus 
only {\it positive} emission from galaxies and stars. The minimum between two images
shifted by 14.9$'$ (in celestial coordinates) eliminates all non-cometary emission on
angular scales smaller than 14.9$'$. There is some diffuse, extended interstellar medium
emission in the field, which is not completely removed, but its brightness is much smaller 
than the comet emission. Also, there are rare cases where, by chance, galaxies overlap between the images at the two epochs; we verified none of the comet fragments are galaxies by
inspecting the single-epoch images for each fragment.
A minor caveat to the method is that {\it noise} is both positive and negative (relative to the
mean), so our method will generate an image with a systematic negative bias. The
noise is primarily due to shot noise from the sky brightness, so the negative bias is
proportional to the  square-root of the sky brightness. The signal-to-noise of the images
is high ($>100$) due to the bright zodiacal background, so the negative bias
is small: $<1$\% estimated from numerical simulation.
It is important to note that the background subtraction required no assumption about the cometary
emission, so cometary emission on all angular scales was preserved.
The most prominent artifacts in the comet-only image are $5'$-wide vertical stripes; these are caused
to gradual detector drifts, and their shape on the image is due to the scanning strategy.

\begin{figure}
\plotone{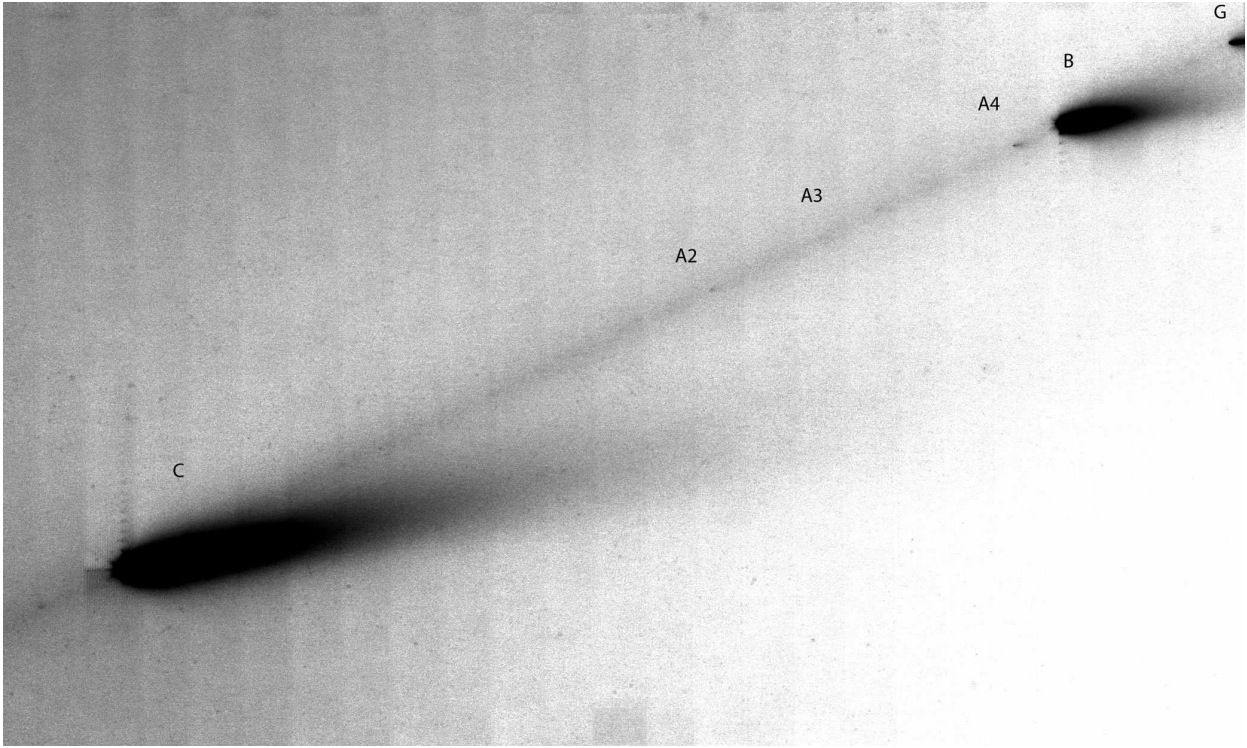}
\figcaption{MIPS 24 $\mu$m image from 2006 Apr 1. Major fragments C, B, and G are labeled, as are the smaller fragments
A2, A3, and A4. North is up, East is to the left, and the image size is $123'\times 83'$.
The image was background-subtracted, using the
minimum between the two observing epochs as described in the text.
\label{mar06min}}
\end{figure}

Figure~\ref{mar06min} shows the MIPS image. The debris trail is clearly evident, spanning the entire image from ahead of fragment C to well behind fragment B. The MIPS arrays
were scanned across the field in 26 fast scan legs with full-array ($5'$) spacing
between legs. A total of 6 hr of observing time was spent on this image.

\begin{figure}
\plotone{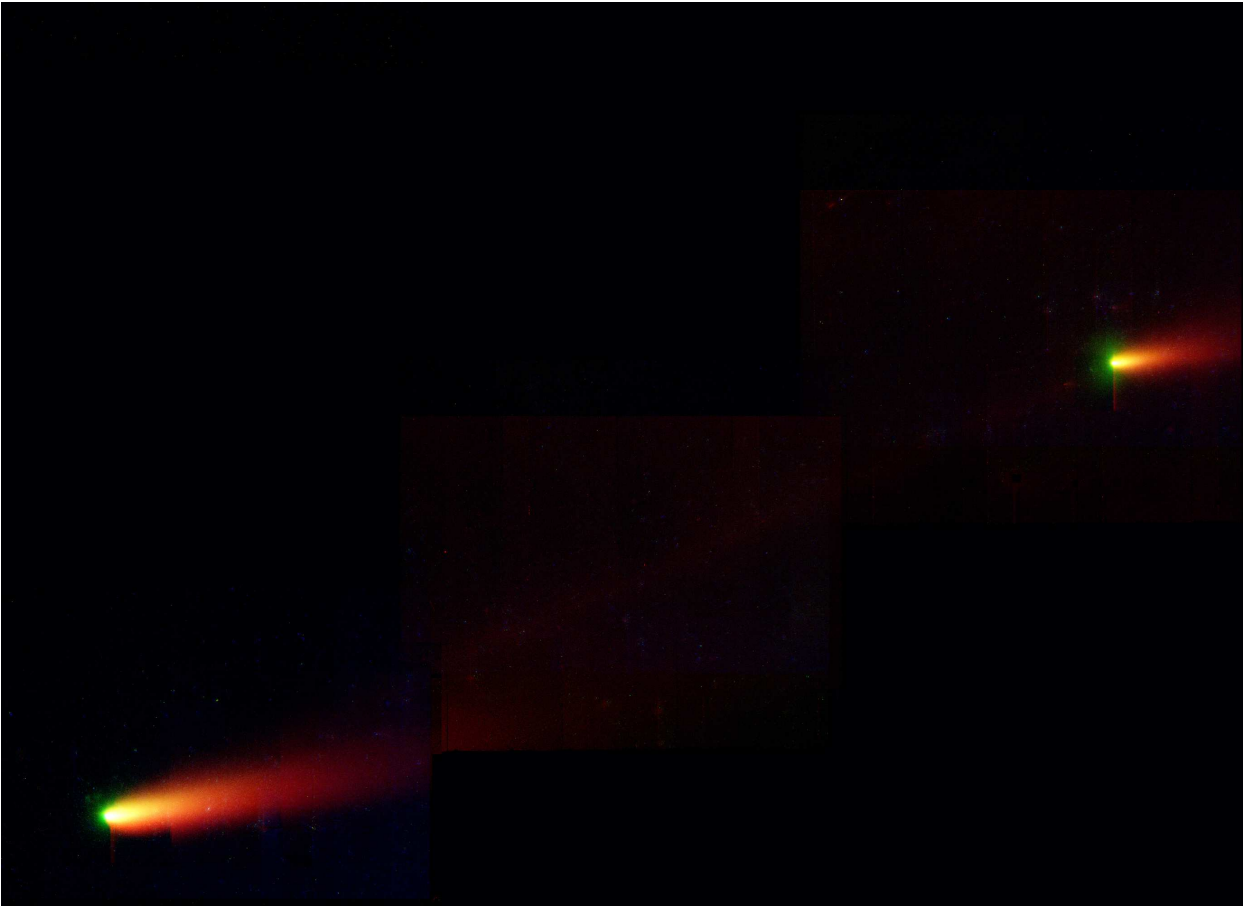}
\figcaption{IRAC color-composite image from 2006 Mar 28. Blue is 3.6 $\mu$m, green is 4.5 $\mu$m, orange is 5.8 $\mu$m, and red is 8 $\mu$m. 
North is up, East is to the left, and the image size is $110'\times 63'$.
The diffuse green emission, extending sunward from
the main fragments, is due to CO and CO$_{2}$ gas
emission (\S\ref{cosection}).
The image was are background-subtracted, using the
minimum between the two observing epochs as described in the text.
\label{monmosirac}}
\end{figure}

Figure~\ref{monmosirac} shows the IRAC image. 
Because the fragment and debris field of the comet spanned such a large field on the sky (exceeding
the maximum observation size), the image was created by 
observing three tiles staggered along the orbit of the comet.
For each tile, a raster map of $10\times 6$ pointings spaced by one 
$5'\times 5'$ IRAC field of view was made, and two 12 sec frames were taken
at each pointing. A total of 4.4 hr observing time was spent on this image.

\subsection{2006 May}
The Director's Discretionary observations (Program ID 274) in 2006 May 
were performed with MIPS and spanned a larger area both on the sky
and in the comet's frame, with excellent viewing conditions---the
comet was at its closest approach to {\it Spitzer} and the orbit plane
angle was $25^{\circ}$. The observations were designed and proposed
immediately after receipt of the 2006 Apr images were received. The field
size for the 2006 May observations was calculated so as to cover
the orbit from well ahead of fragment C to behind the 
by-then-discovered fragment G (which was near the edge of the 2006 Apr
observation), after adjusting for the relative distance between the comet
and the observatory on the two dates.
It is worth noting that the viewing geometry from {\it Spitzer} was somewhat
different from that for ground-based observations. 
As shown in Figure~\ref{obsfig},
the comet was much closer to Earth (0.10 AU) than it was to {\it Spitzer} (0.37 AU); in fact, 
the comet was closer to Earth than {\it Spitzer} was to Earth (0.32 AU).

\begin{figure}
\plotone{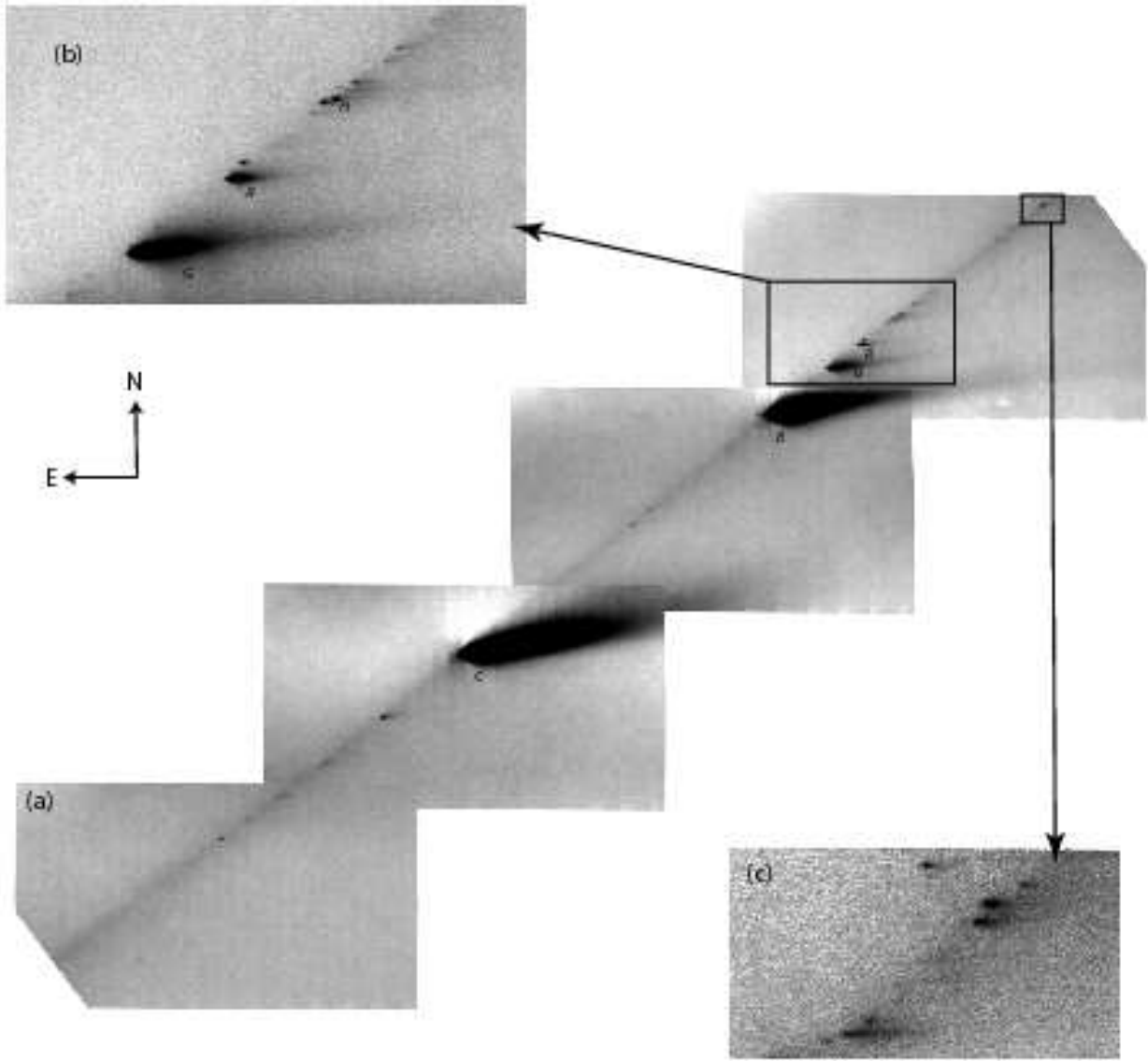}
\figcaption{MIPS image from 2006 May 4--6. These images are background-subtracted, using the
minimum between the two observing epochs as described in the text. 
The orientation is indicated with arrows pointing N and E, and the length of those arrows is 
$17^\prime$. Some major fragments are labeled.
Panel (a) is the is the
entire mosaic, combining the 4 large tiles. 
Panels (b) zooms into a region behind 73P-G that is  particularly  rich in fragments.
Panel (c) zooms into the portion of the observation furthest behind the main fragments,
and it includes the fragment located furthest perpendicular to the project orbit.
\label{monmos}}
\end{figure}

Figure~\ref{monmos} shows the MIPS image.
To cover the large field (larger than that which can be covered in a single
scan map) the field was broken into 4 rectangular tiles, with centers
aligned along the projected orbit including overlaps at the corners.
Each tile is the same type of observation as the 2006 Apr image.
A total of 24 hr of observing time was spent on this image.

\subsection{2007 Jan}
The {\it Spitzer} Cycle 3 Guaranteed Time observations (Program ID 30010) in
2007 Jan were designed to cover a similar portion of the orbit, but at much less
favorable viewing geometry. As shown in Figure~\ref{obsfig}, the comet was much more distant;
further, the orbit
plane angle was only $2.3^{\circ}$, making the trajectories
of small and large particles overlap on the sky. 
The observation was made using the MIPS 24 $\mu$m photometry mode, with one cycle of 10-sec-exposure photometry performed at each of a set of cluster positions spaced along the projected orbit.

\begin{figure}
\plotone{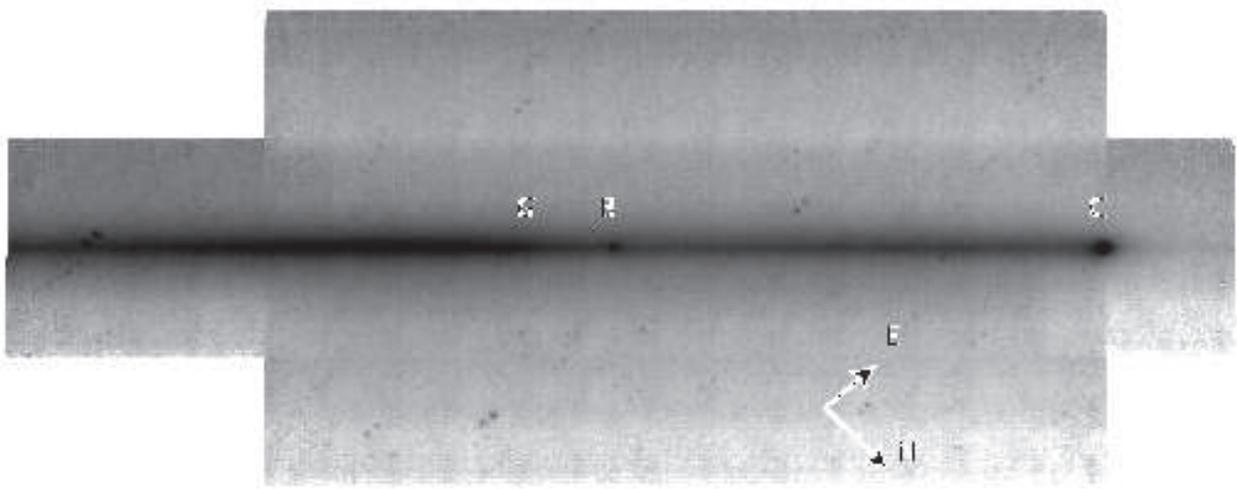}
\figcaption{MIPS image from 2007 Jan. The image is rotated for compact display; the orientation of celestial North and East are indicated.
The image covers $46'\times 18'$. Fragments C and B are labeled, as is the expected location of
fragment G if it had remained at the same spacing behind fragment B, relative to the separation
between fragments B and C, in 2007 Jan as it had in 2006 Apr and 2006 May.
\label{sw3jan07}}
\end{figure}

Figure~\ref{sw3jan07} shows the MIPS image.
The anti-solar and projected
comet motion vectors are nearly aligned on the sky, so the tail 
(small particles from the current orbit)
and
trail or meteoroid stream
\citep[larger particles from previous orbits, cf.][]{reachtrail}
overlap the fragment distribution.
Though the heliocentric distance was 2.8 AU, and the sublimation rate of H$_2$O ice should be relatively low,
the large and small particles cannot be readily disentangled, as they could in the 2006 Apr-May observations.

Only fragments C and B are detected, despite having enough sensitivity to have detected fragment G if it scaled similarly to the other main fragments. 
The lack of a detectable condensation at the predicted location of fragment G suggests this fragment completely disrupted.
This is consistent with the appearance of fragment G in {\it Hubble} Space Telescope images: instead of a dominant nucleus
with retinue of smaller bodies, fragment G appeared as a swarm of smaller bodies.

The diffuse emission in 2007 Jan closely follows the projected orbit. 
A debris trail stretches from the one edge of the image to the other, i.e. from well ahead of fragment C to well behind fragment B. 
The emission is far brighter than a typical debris trail at 2.8 AU from the Sun. 
Ahead of fragment C, the diffuse emission is at 0.1 MJy~sr$^{-1}$, which is typical of debris trails produced naturally
by comets and representing mass loss on previous revolutions around the Sun  \citep[cf.][]{reachtrail}.
But following fragments C and
especially B, the brightness is much higher and has an atypical brightness distribution, as we will discuss further in \S\ref{sectrail} below.

\clearpage

\section{Properties of fragments}

\subsection{Spectral energy distributions of the comae of fragments B and C}

\begin{figure}
\plotone{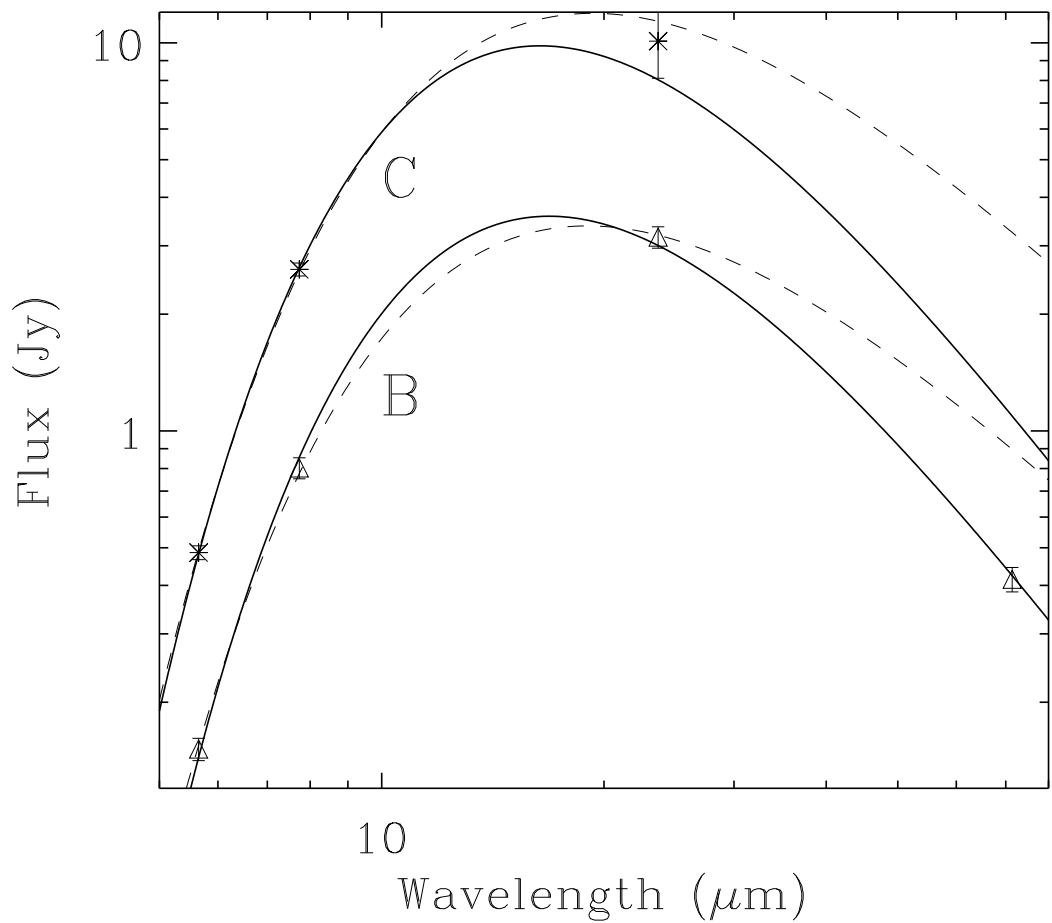}
\figcaption{Spectral energy distribution of fragments B and C 
based on observations on 2006 Mar 28 and Apr 1. 
The dashed lines for both fragments show a fit with $\alpha=0$
and free temperature, fitted only to the 5.8--24 $\mu$m 
photometry. These fits clearly overpredict the long-wavelength flux.
For fragment B the thick solid line shows a best fit
with a free temperature and emissivity index (see text for definition), with best-fit values
$\alpha=0.79\pm 0.12$ and $T=236\pm 8$ K. 
For fragment C the thick solid line shows the
best fit with fixed $\alpha$ the same as found
for fragment B, and
free temperature $T=239\pm 4$.
\label{aprtemp}}
\end{figure}

Figure~\ref{aprtemp} and Table~\ref{aprflux} show the spectral energy distribution of
fragments B and C. Fluxes were measured within a $25''$ radius aperture,
and they are dominated by dust in the coma.
Fragment B was measured at all wavelengths (3.6--70 $\mu$m), but
fragment C was saturated at 24 $\mu$m and
was not covered by the operative portion of the 70 $\mu$m
detectors.
To estimate the 24 $\mu$m flux of fragment C, we replaced the central 3-pixel 
($7.5''$) radius
region centered on the nucleus (the only pixels that are saturated) with 
a $1/\rho$ brightness distribution, convolved with a $5''$ gaussian FWHM beam
and fitted to the coma from 10--30$''$ from the nucleus. The estimated flux in the saturated
portion of the image is a small fraction ($<10$\%) of the total; nonetheless, we adopt a 
conservative 40\% uncertainty. Note that the observations at wavelengths shorter than
10 $\mu$m were taken on 2006 Mar 28, while those at longer wavelengths were taken 
on 2006 Apr 1. Fragments B and C were brightening through 2006 Mar and Apr. 
We fit greybody, $F_\nu \propto B_\nu(T)$, and 
modified greybody $F_{\nu}\propto \nu^{\alpha} B_{\nu}(T)$,
where $T$ is the temperature and $B_{\nu}$ is the Planck function, to the
spectral energy distributions, allowing for a fractional brightening over the 4-day
interval, between 2006 Mar 28 and 2006 Apr 1, in the range 1--1.4. 
If the brightness increase is more than 1.4, then it is not possible to put a
modified blackbody through the 5.8--24 $\mu$m data points that are
completely dominated by thermal dust emission.

The spectral energy distribution of fragment B could not be adequately fit using a 
greybody spectrum, irregardless of any possible brightness variation between
the two observing dates.
The observed 70/24 $\mu$m flux ratio of $0.13\pm 0.02$, on 2006 Apr 1, requires
a color temperature $> 800$ K, which is far higher than comet grains
at 1.33 AU are expected to be, and which is completely inconsistent with the
color temperature from of $255\pm 12$ K measured from the 5.8--8 $\mu$m brightness
on 2006 Mar 28.
A modified blackbody yields an acceptable fit to the spectrum, as long as the
emissivity of the grains decreases as the $-0.79\pm 0.12$ power of the wavelength.
The grain temperature from such a fit is $236\pm 8$ K, which is consistent with
the 238 K expected for rapidly-rotating isothermal grains at 1.37 AU from the Sun.

The depressed 70 $\mu$m brightness is likely due a dominant contribution from grains too small
to emit efficiently at such long wavelengths. Particles smaller than $\lambda/2\pi=11$ $\mu$m are
expected to be in the small particle limit, where their emissivity will decrease as $\lambda^{-1}$.
The observed emissivity index of 0.79 suggests the dominant particles are in the range 2 to 20 $\mu$m in radius.

\begin{deluxetable}{lll}
\tablecaption{Fluxes of fragments B and C on 2006 Mar 28\tablenotemark{a}\label{aprflux}}
\tablehead{
\colhead{Wavelength} & \colhead{Fragment C Flux (Jy)} & \colhead{Fragment B Flux (Jy)} }
\startdata
5.65 & $0.37\pm 0.02$ & $0.115\pm 0.01$\\ 
7.74 & $2.1\pm 0.1$   & $0.63\pm 0.05$ \\ 
23.7 & $9.6\pm 2$ & $3.0\pm 0.2$ \\
71.4 & ... & $0.40\pm 0.03$\\
\enddata
\tablenotetext{a}{Fluxes are measured within a circular 25$''$ radius aperture. The 23.7 and 81.4 $\mu$m fluxes were scaled from 2006 Apr 1 measurements to match the 2006 Mar 28 
observations at the other wavelengths.}
\end{deluxetable}

\clearpage

\subsection{Locations and fluxes of intermediate-sized fragments}

\def\extra{\tt\bf ??? NOTES:
measure fluxes, radii, comaAfrho and compare to individual comets

CML:  For the radii, should we list the upper limits to the estimated size using the total flux as observed, or our best guess as to the coma-removed nuclear flux + derived radius?

JV: I would be inclined to go for our best guess + derived radius

MVS: upper limits for radii
}

Table~\ref{fragtab} lists the fragments detected in the 24 $\mu$m images at each of the three observing epochs. Only the primary fragments B and C, which were most likely produced in
the main 1995 splitting event, are present in all three epochs. Fragment G was very bright before the 2006 perihelion passage but was not recovered afterward. Our 2007 Jan observation shows no nucleus or coma at the location of fragment G. The 2006 May observation shows numerous fragments, as did ground-based optical observations around the same time. As discussed later in this paper, the infrared and optical fragments could not be readily matched. We used the
JPL Horizons utility to calculate locations of all fragments for which orbits had been calculated (based on ground-based optical astrometry). Fragment G corresponds well with the third-brightest mid-infrared fragment (2006 Apr and May), and the optically-bright
fragment R's predicted
location is close to an infrared-bright fragment in 2007 May. These identifications are
noted in Table~\ref{fragtab}. Three other fragments whose locations in the Spitzer image
are close to predicted locations of fragments with calculated orbits are listed with a question mark in Table~\ref{fragtab}. The remaining fragments have no clear counterpart.
We will refer to fragments in the 2006 Apr image as A1--A5, and those in the 2006 May
image as S1--S55, except for the well-identified fragments. It is likely that
fragment A4 and S16 (located just ahead of fragment B) are the same object.

The flux of each fragment was measured using annular-background-subtracted aperture photometry.
Fragments B and C are saturated in the images so their fluxes could not be measured.
Table~\ref{fragtab} lists the fluxes as well as a note on the presence or lack of a dust tail,
by which we mean diffuse emission that extends from the fragment in the anti-solar direction.
The observed flux of a fragment combines information about the nucleus size and the current dust production rate. 
For bare nuclei, the flux at 24 $\mu$m is $F_{1} D^{2}$ where $F_{1}$ is the flux of 
a body with a diameter of 1 km.
We calculated $F_{1}=88$ mJy for the asteroid standard thermal model \citep{lebofsky86},
90 mJy for a rapidly-rotating (isothermal-latitude) model, and
42 mJy for the near-Earth-asteroid thermal model \citep{neatm},
assuming albedo 4\% and  the 2006 May observing geometry from {\it Spitzer}.
For these calculations it is important to note the high phase angle 
(Sun-Comet-{\it Spitzer}) of
$72^\circ$ is outside the normal range for main belt asteroids.
We adopt $F_1=50$ mJy for diameter estimates this paper.
Assuming an albedo of 0.04, 
the size of fragment C was measured using {\it Hubble} Space Telescope
to be $0.68\pm 0.04$ km \citep{toth05}; the nuclear flux at 24 $\mu$m would then be 23 mJy. 
The observed 24 $\mu$m flux is 2 orders of magnitude higher than this value, suggesting it is dominated by thermal emission
from dust surrounding the nucleus.

Many of the smaller fragments have tails, showing they, too, are actively producing dust. 
The observed fluxes yield only upper limits to the fragment sizes;
the upper limits can be obtained from the fluxes, $F$, in Table~\ref{fragtab}
as $D < (F/50\,{\rm mJy})^{0.5}$ km.
The faintest fragments,
with fluxes $\sim 3$ mJy have a nuclear diameter limits $<250$ m. 
These limits are likely much
larger than the actual fragment sizes, since the fragments are active and
their mid-infrared fluxes are much smaller than that of fragments C and B.

\begin{figure}
\epsscale{.8}
\plotone{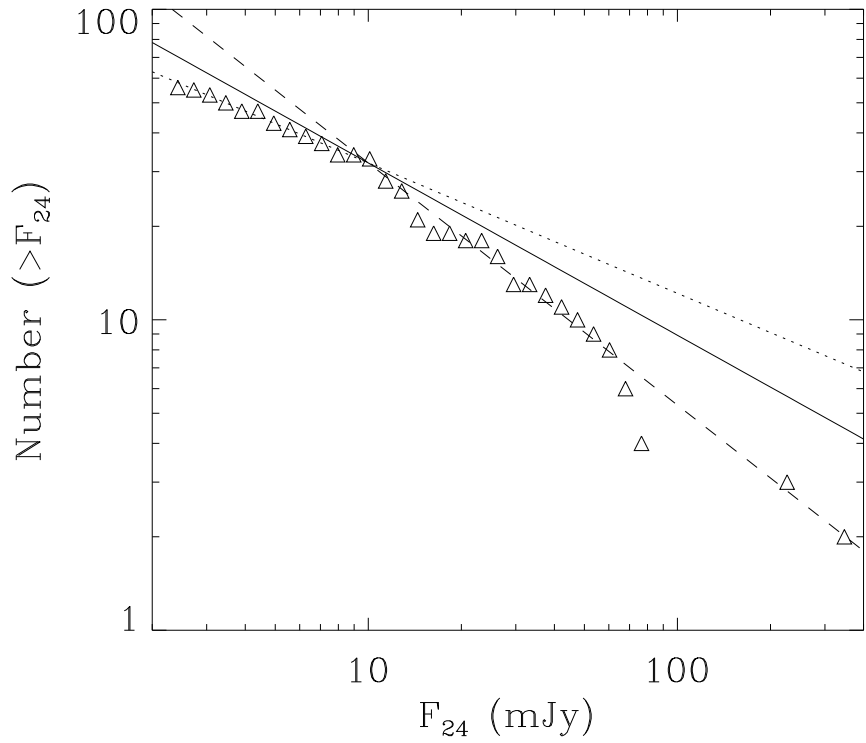}
\epsscale{1}
\figcaption{The cumulative 24 $\mu$m flux distribution of 56
  fragments observed in May 2006.  
Three power-law fits to the  luminosity distribution are presented:
the solid line shows the slope derived from Subaru observations by \citet{fuse};
the dotted line is a fit for fluxes less than 10~mJy;
and the dashed line is a fit for fluxes larger than 10 mJy.
\label{sizedist}}
\end{figure}

The distribution of observed fluxes gives some indication of the size
distribution, if the dust production rate scales with nuclear size.
In fact the dust production is not expected to be perfectly correlated
with size and may even be episodic, so keep in mind that
the  following analysis of
fragment `size' distribution may be skewed relative to the
nucleus size distribution. However, the fluxes
(from aperture photometry) measured in the infrared images should
be directly proportional to those measured from optical
images, so that we can compare the {\it Spitzer} results to previous optical measurements.
Figure~\ref{sizedist} shows the cumulative luminosity distribution of
the 53 fragments with fluxes measured 2006 May.  We fit pieces of the
cumulative luminosity distribution with a power-law function to
approximate the true cumulative luminosity function.  Our results are
presented in Table~\ref{table:clf}.  The cumulative luminosity
function has a clear break in the power-law slope at a flux of 10 mJy.

The slope for fragments fainter than 10 mJy (-0.42) is close to the
slope found by \citet{fuse} (-0.555) in their Subaru observations of
mini-fragments near 73P-B.  The mini-fragments near 73P-B fall along
its tail and are likely icy (as discussed in \S\ref{dynamicssection}); 
the Spitzer-detected fragments fall along
the comet's orbit and likely have less exposed ice, as explained in
detail in \S\ref{dynamicssection} below.  The Spitzer-detected fragments are also
likely somewhat larger ($\sim50$~m) than the Subaru-detected
mini-fragments ($\sim20$~m). 
The similarily of flux distribution slopes for the fragments along the trail
(from the breakup in 1995) and the mini-fragments in 73P-B's tail (from the
breakup of 73P-B in 2006) suggests that the fragmentation process itself 
operates similarly over a wide range of size scales.

If we interpret the luminosity distribution slopes as nuclear size distributions,
where the
size distribution is
\begin{equation}
  n(R)dr \propto R^{-q}dr,
\end{equation}
then the inferred size distributions slopes are 
$q=1.84$ for small ($F<10$ mJy) fragments and
$q=2.56$ for large ($F>10$ mJy) fragments.
We compare our derived size distributions to other comet size
distributions in Table~\ref{table:sized}.  
For a system in collisional equilibrium, $q=3.5$, which
is much steeper (i.e. more, smaller particles) 
than we observe and steeper than any of the comet size
distributions in Table~\ref{table:sized}.
Our large fragment size
distribution is similar to that derived from three different studies: 
(1) the size distribution of fragments
discovered after the catastrophic breakup of comet C/1999~S4 (LINEAR),
$q=2.72\pm0.60$ \citep{makinen01}, 
(2) the inferred initial size
distribution of Kreutz group sungrazing comets
\citep[$R=20-200$~m;][]{sekanina03}, and 
(3) the observed size
distribution of ecliptic comet nuclei for $R>1.6$~km \citep{LamyNuc}.

These agreements suggest that 
the break in the slope at $F_{24}=10$~mJy indicates a
change in the size distribution slope, rather than a change in the
dependence of coma brightness on fragment size.
The change in size distribution slope doe snot necessarily reflect
the fragmentation process; instead,
the relative deficit of smaller fragments may indicate that they have a short lifetime.
Fragments with a short lifetime will be more abundant in images taken soon after 
an event that produces small fragments. They will also be more abundant in 
higher-angular-resolution images
taken close to the site of their production.
Both of these conditions are met for the Subaru and HST observations, which
detect more of the very small fragments than we could with {\it Spitzer}.

\begin{deluxetable}{lrrrrl}
\tablecaption{Fluxes and locations of Fragments of 73P/Schwassmann-Wachmann 3\label{fragtab}}
\tablehead{
\colhead{Fragment} & \colhead{$F_{24}$} & \colhead{$D_{xC}/D_{BC}$\tablenotemark{a}} & 
\colhead{$D_{xC}/\Delta$\tablenotemark{b}} & \colhead{$z/\Delta$\tablenotemark{c}} & \colhead{Comment}
\\&(mJy)&&($^\prime$)&($^\prime$)
}
\startdata
\cutinhead{2006 Apr}
C	& 4760	& 0		& 0     & 0      & tail$>70^\prime$\\
A2	& 4.5	& -0.63	& -61.3 & $<0.1$ & \\ 
A3	& 1.9	& -0.81	& -79.1 & $<0.1$ & fuzz\\
A4	& 10.8	& -0.95	& -93.3 & $<0.1$ & fuzz\\
B	& 2690	& -1		& -98.2 & $<0.1$ & tail$>18^\prime$\\
G	& 138	& -1.18	& -115.8& $<0.1$ & coma+tail\\
\cutinhead{2006 May}
 S1 &    2.3 &   1.11 &   149.1 &  2.23 & fuzz \\
 S2 &    4.7 &   0.96 &   129.1 &  1.93 & fuzz \\
BR? &   58.9 &   0.79 &   105.7 &  1.25 & tail $3^\prime$ \\
AT? &    4.3 &   0.71 &    95.1 &  0.79 & \\
 S5 &   12.9 &   0.60 &    81.0 &  0.76 & tail $3^\prime$ \\
 S6 &    5.2 &   0.48 &    64.8 &  0.24 &  \\
AV? &   22.7 &   0.47 &    62.8 &  0.08 & tail $3^\prime$\\
 S8 &    3.6 &   0.45 &    61.1 &  0.31 &  \\
 S9 &   70.1 &   0.27 &    36.8 &  0.32 & tail $4^\prime$ \\
S10 &   25.2 &   0.26 &    35.3 &  0.25 & tail $3^\prime$ \\
  C &   ...    &   0.00 &     0.0 &  0.00 & saturated \\
S12 &   38.1 &  -0.55 &   -74.0 & -0.40 & tail $2^\prime$ \\
S13 &   13.8 &  -0.60 &   -80.8 &  0.24 &  tail $2^\prime$\\
S14 &   10.5 &  -0.62 &   -82.7 & -0.08 &  tail $2^\prime$\\
S15 &    4.7 &  -0.86 &  -116.0 &  0.16 &  \\
S16 &    3.2 &  -0.95 &  -127.7 & -0.08 & A4? tail $1.5^\prime$ \\
  B &  ...   &  -1.00 &  -134.3 &  0.28 & saturated; tail $90^\prime$ \\
  G &  336.9 &  -1.19 &  -159.5 &  0.21 & tail $30^\prime$ \\
S19 &   13.7 &  -1.22 &  -163.6 &  0.59 & fuzz \\
  R &  219.3 &  -1.29 &  -172.7 & -0.10 & tail $5^\prime$ \\
S21 &   47.8 &  -1.30 &  -174.7 &  0.57 & BN? coma \\
S22 &    2.9 &  -1.33 &  -178.2 & -0.00 & fuzz \\
S23 &    2.9 &  -1.34 &  -179.8 & -0.13 & fuzz \\
S24 &    5.7 &  -1.36 &  -183.3 &  0.40 &  \\
S25 &    5.7 &  -1.37 &  -183.5 &  0.15 &  \\
S26 &    3.3 &  -1.37 &  -184.2 &  0.95 & \\
S27 &   55.5 &  -1.38 &  -185.1 &  0.49 & N??; tail $2^\prime$ \\
H   &   64.4 &  -1.39 &  -186.3 & -0.05 & tail $2^\prime$ \\
S29 &   13.7 &  -1.39 &  -187.4 &  0.67 & BQ??\\
S30 &    3.7 &  -1.40 &  -187.7 & -0.03 &  \\
S31 &   27.6 &  -1.41 &  -188.9 &  0.15 & BP??\\
S32 &    9.9 &  -1.43 &  -192.7 & -0.01 &  \\
S33 &   10.5 &  -1.44 &  -193.1 &  0.16 &  \\
S34 &   15.7 &  -1.45 &  -194.6 &  0.33 &  \\
S35 &    6.9 &  -1.48 &  -198.9 &  0.67 &  \\
S36 &    6.4 &  -1.49 &  -199.8 &  0.13 &  \\
S37 &    6.8 &  -1.54 &  -206.6 &  0.25 &  \\
S38 &    3.5 &  -1.59 &  -213.6 &  0.02 &  \\
S39 &   11.4 &  -1.67 &  -223.9 &  0.38 & compact \\
S40 &   10.7 &  -1.70 &  -228.6 &  0.29 &  \\
S41 &    4.6 &  -1.72 &  -231.5 &  0.54 &  \\
S42 &   34.2 &  -1.74 &  -234.0 &  0.82 & tail $1^\prime$ \\
S43 &    4.8 &  -1.76 &  -236.7 &  0.11 &  \\
S44 &   19.1 &  -1.77 &  -237.9 & -0.38 & tail $1^\prime$ \\
S45 &   11.0 &  -1.78 &  -239.3 & -0.53 &  \\
S46 &    3.2 &  -1.80 &  -241.5 &  0.04 &  \\
S47 &   28.0 &  -1.83 &  -246.5 &  4.16 & tail $1^\prime$ \\
S48 &   26.4 &  -1.84 &  -246.9 &  0.14 &  \\
S49 &   43.7 &  -1.85 &  -247.9 &  0.56 & tail $1^\prime$ \\
S50 &   12.4 &  -1.86 &  -250.1 &  0.19 & \\
S51 &    8.8 &  -1.88 &  -252.3 &  0.83 &  \\
S52 &    7.5 &  -1.91 &  -256.3 &  0.30 & \\
S53 &    7.3 &  -1.96 &  -263.8 &  0.29 & \\
S54 &   14.9 &  -2.05 &  -275.5 &  0.41 & tail $1^\prime$ \\
S55 &   10.8 &  -2.05 &  -275.8 & -0.08 & \\
\cutinhead{2007 Jan}
C	& 55.1	& 0		& 0 \\
B	& 13.2	& -1		& -18.1\\
{\it G}	& $<$0.2 	& -1.18\tablenotemark{d}	& ...\\
\enddata
\tablenotetext{a}{Separation of fragment from 73P-C, in units of the separation between fragments B and C}
\tablenotetext{b}{Separation of fragment from 73P-C, in arcmin}
\tablenotetext{c}{Perpendicular displacement of fragment from the projected orbital path of 73P-C, 
divided by the Comet-Observer separation, in arcmin}
\tablenotetext{d}{assumed location of fragment, based on previous observations}
\end{deluxetable}

\clearpage

\begin{deluxetable}{lll}
\tablecaption{Best-fit power-law slope of the fragment cumulative
  luminosity function in May 2006.\label{table:clf}}
\tablehead{
\colhead{Flux Range (mJy)} & \colhead{$k$\tablenotemark{a}} &
\colhead{$\sigma_{MAD}$ (dex)}\tablenotemark{b}
}
\startdata
  $F>10$ mJy    & -0.78 & 0.04 \\
  $F<10$ mJy    & -0.42 & 0.02 \\
\enddata
\tablenotetext{a}{Power-law slope determined by the least absolute
  deviation method.}
\tablenotetext{b}{Mean-absolute deviation in log space.}
\end{deluxetable}

\begin{deluxetable}{lll}
\tablecaption{Comet fragment differential size distributions. \label{table:sized}}
\tablehead{
\colhead{Comet} & \colhead{$q$} & \colhead{Source} }
\startdata
  73P (small)         & 1.84                            & this work \\
  73P (large)         & 2.56                            & this work \\
  73P-B tail fragments & 2.11                      & \citealt{fuse} \\
  C/1999 S4 (LINEAR)  & $2.72\pm0.60$\tablenotemark{a}  & \citealt{makinen01} \\
  Kreutz group        & 2.7--3.0\tablenotemark{b}       & \citealt{sekanina03} \\
  \hline
  All comets          & $2.9\pm0.3$                     & \citealt{LamyNuc} 
\enddata
\tablenotetext{a}{Inferred from the fragment water production rates.}
\tablenotetext{b}{Derived from the cumulative mass distribution slope
  range, -0.56 to -0.68.}
\end{deluxetable}

\clearpage

\section{Distribution of fragments}

Determining the identity and location of each of the fragments in our infrared image series is a non-trivial exercise. 
The location of each fragment was rotated into a coordinate system measuring separation from fragment C along the line directed toward fragment B, and the distance perpendicular to this line. To remove the geometric projection due to changing distance from the telescope, the distance along the orbit was divided by the separation between B and C. There are 10 fragments leading C, 5 fragments between B and C, and 38 fragments following B (for a total of 55 fragments). 
The infrared fragments span nearly the entire observed field, ranging from
-2.07 to +1.13 times
$\vec{BC}$, where  $\vec{BC}$ is the vector from B to C with origin at C.
For comparison, orbits of 66 fragments have been generated by the Minor Planet Center and 
JPL/Solar System Dynamics Group. The similarity of the number of infrared and optical fragments suggests there may be good correspondence, and most of the infrared fragments may have optical counterparts. 


In an effort to cross-identify the fragments, we generated ephemerides using the
Horizons service for all 66 fragments of 73P with known orbits. 
Figure~\ref{allfrags} shows the distribution of fragments along the orbit of 73P in May 2006. 
There are 13 ephemeris positions ahead of C, 7 between C and B, and 44 behind B, 
The positions range from -2.95 to 1.05 times the separation between B and C, relative to C (except one fragment, S, predicted to be at 6.6 times further behind C than B).
The distribution of predicted positions is very similar to the distribution of infrared fragments, 
especially after removing 7 positions that fall outside the infrared image behind B.
Good matches were found for the major fragments C, B, G, and R.
Beyond that, to our surprise, few convincing matches were found. 
Likely matches include fragments BR and AT, for which the position coincidence is
much better than expected by chance. Plausible matches include optically-bright
M and N with moderately bright infrared fragments, though the deviations are large and multiple matches are possible.
Inspecting Figure~\ref{allfrags} shows that part of the reason for the lack of
matches is that the distance of the fragments perpendicular to the fragment C orbit 
is very different for the observed ({\it Spitzer}) 
and ephemeris positions. The root-mean-square perpendicular separation is $0.6'$ for the observed positions and $10'$ for the ephemeris positions. 
Typical quoted uncertainties in the predicted ephemeris positions are $<0.1'$.

Even if we ignore the out-of-plane motion, the locations of predicted positions along the orbit are different from those in the infrared image in detail. For each infrared fragment, we found the closest optical fragment along the orbit (neglecting perpendicular distance). The separation between the infrared and closest optical fragments has a distribution with a core dispersion of 0.005 $|\vec{BC}|$, containing $\sim 30$ fragments, with a wider distribution of width 0.03 $|\vec{BC}|$.
The core of this distribution could contain a significant number of matching objects, though its dispersion ($40''$ on the sky) is broader than the 
resolution ($5''$).
Fragments C, B, G, and R have infrared-optical matches within $5''$. 
Using a set of random fragment locations, the wider part of the separation distribution is reproduced. We have noted in 
Table~\ref{fragtab} those fragments with optical matches within $5''$ as
identifications in the first column; those within $10''$ as identifications with
``?'' in the first column; and those within $40''$ in the
comment field with ``?''.
Figure~\ref{allfragsapr} shows that the situation in 2006 Apr 1 is similar to that in 2006 May 5: apart from the brightest fragments C, B, and G, there are no matches between the predicted positions and the other infrared fragments. 

\begin{figure}
\epsscale{.9}
\plotone{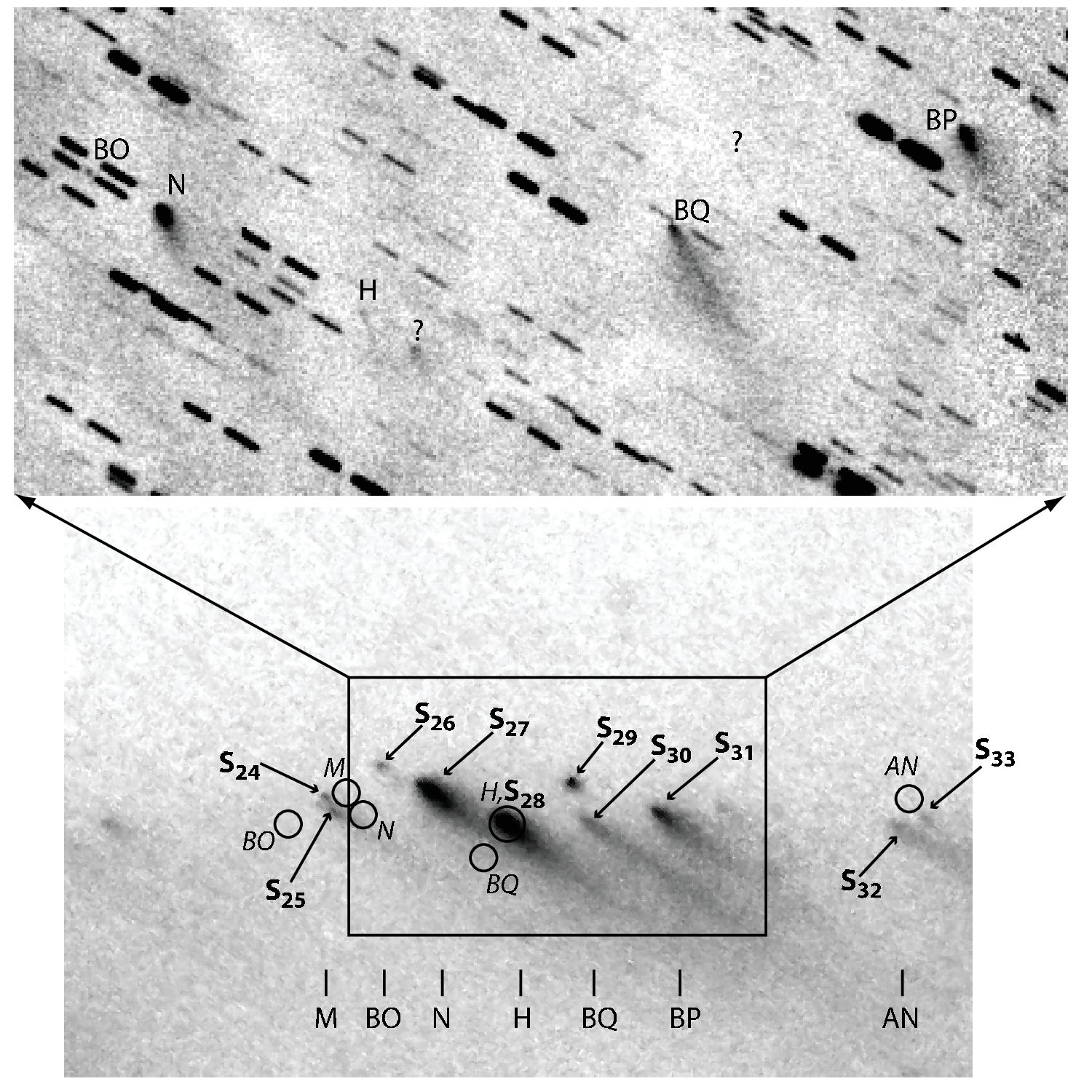}
\epsscale{1}
\caption{\small Zoom into the region around fragments 73P-M, N, and H.
The top panel is an optical image 
(courtesy Peter Birtwhistle, Great Shefford Observatory West Berkshire, U.K.) taken 2006 Apr 30 covering a $15'\times 7'$
field of view with the identified fragments labeled in accordance
with the astrometry reported to the Minor Planet Center.
The bottom panel is a section of the {\it Spitzer} image taken
on 2006 May 4 (during 12:49 to 15:50 UT),
with a box
showing the outline of the optical image of the top panel.
Bold labels of the form {\bf S}$_{n}$ with arrows correspond 
to the fragment list in Tab.~\ref{fragtab}.
Circles with italicized labels indicate positions of fragments as
predicted positions of fragments from ephemerides calculated by
Horizons using orbit solutions based on optical, Earth-based astrometry.
Vertical bars with labels at the bottom of the bottom panel are
the approximate locations of fragments determined
by P. Birtwhistle, based on their spacing in the optical image.
\label{birtfrag}}
\end{figure}

\begin{figure}
\epsscale{.8}
\plotone{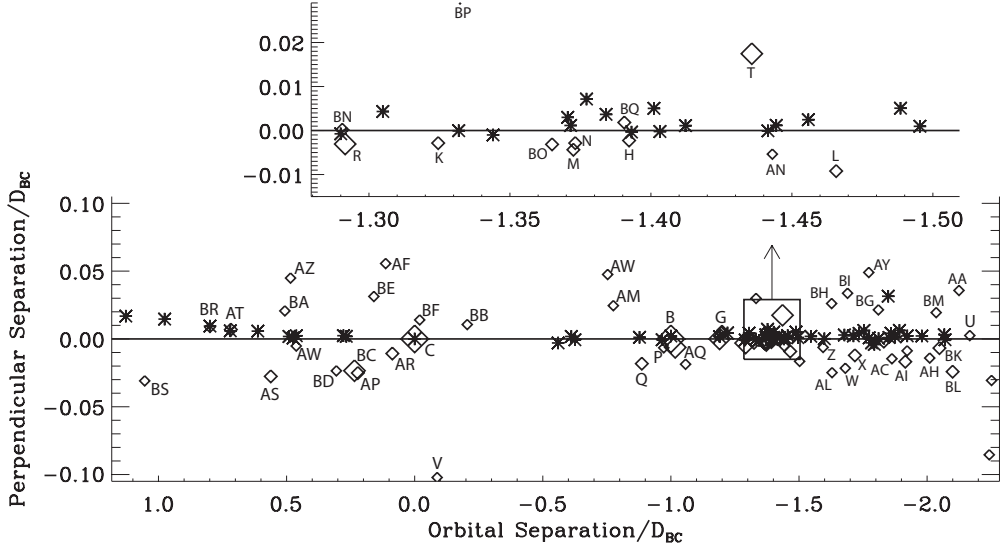}
\epsscale{1}
\caption{Locations of fragments of 73P on 5 May 2006 as predicted using the JPL Horizons ephemerides (diamonds) and as detected with the Spitzer Space Telescope (asterisks). The horizontal axis indicates the distance behind fragment C along the line connecting fragments B and C, in units of the separation between B and C ($D_{BC}$); the vertical axis indicates the distance of the fragment perpendicular to the line running through B and C, also in units of $D_{BC}$.
The size of the diamonds depends on the magnitude from Horizons:
those with no magnitude are smallest, those with magnitudes (H mag) greater than 20 are next, those with magnitudes less than 20 are next, and fragments B and C are shown with large symbols.
The inset shows a zoom into a region particularly rich in fragments.
The lack of correspondence between predicted and observed positions is evident, with the predicted fragments much more scattered perpendicular to the orbit than the observed fragments.
The first astrometric observations included in the orbit calculations for all
fragments was before the date of this {\it Spitzer} image. 
However, nearly all of the fragments were seen only briefly, and their {\it latest}
reported astrometric observations included in the orbit determination were
before the date of the {\it Spitzer} image.
The fragments with astrometry reported after 2006 May 4 are
B, C, H, M, X, AI, AQ, BC, BN, BO, BP, and BR.
A further 11 had astrometry up until 2006 May 1:
G, L, N, R, AC, AS, BF, BI, BK, BQ, and BS.
\label{allfrags}}
\end{figure}

\begin{figure}
\epsscale{.9}
\plotone{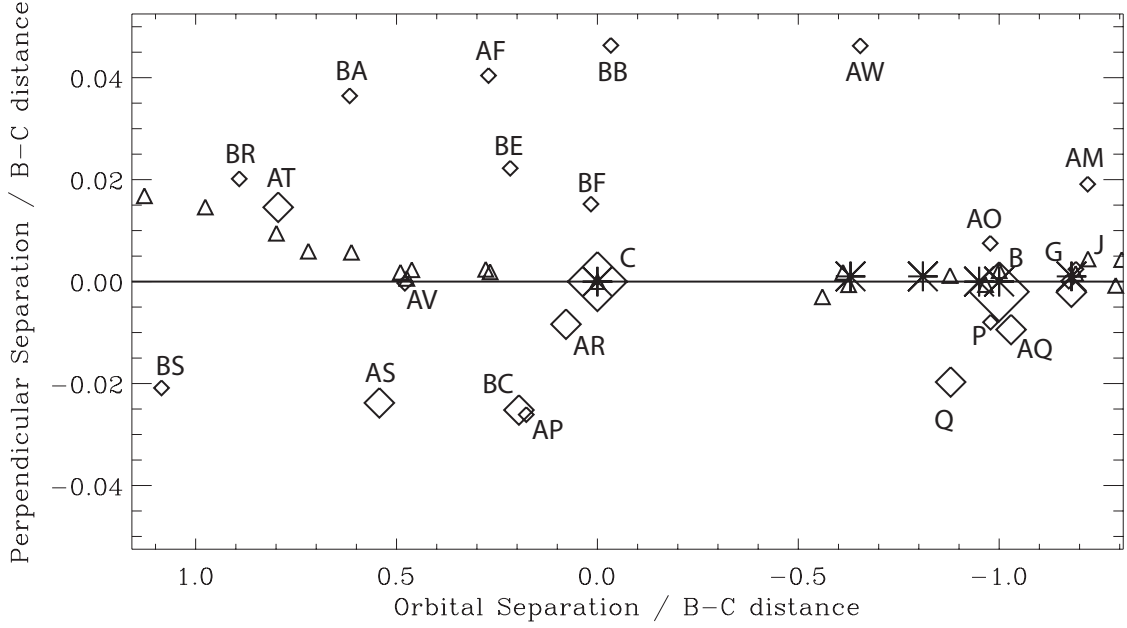}
\epsscale{1}
\caption{Locations of fragments of 73P on 1 Apr 2006 as predicted using the JPL Horizons ephemerides (diamonds) and as detected with the Spitzer Space Telescope (asterisks). Only the positions of the very brightest fragments C, B, and G have corresponding infrared detections. The two fainter, infrared fragments,
located along the orbit between B and C, do not match ephemeris-predicted positions of any fragments.
Note that on 1 Apr, many of the fragments had not yet been optically reported
(so they may not have existed or been activated). The fragments with
astrometry reported before Apr 1 are 
B, C, H, J, K, L, N, M, V, U, P, Q, R, S, W, T, X, Y, and AU.
Another 9 fragments were reported on 2 Apr,
just a day after the first {\it Spitzer} image; these include
Z, AA, AB, AC, AD, AE, AF, AG, and AV.
The other 36 ephemeris positions plotted as diamonds are extrapolations back in
time before the first reported position, so they may not be valid.
\label{allfragsapr}}
\end{figure}

An attempt to match the optical and infrared fragments, using
a procedure very similar to the one described above, was 
undertaken by P. Birtwhistle (website and personal communication). 
One difference between our procedures
was that, instead of using the Horizons-predicted ephemerides,
Birtwhistle sorted fragments by time of perihelion and compared
the {\it Spitzer} image to close-in-time optical images and
astrometry reported to the Minor Planet Center. In general,
finding matches was extremely difficult, as we reported above. 
But there were some sections of the fragment train where matches
appeared plausible. Figure~\ref{birtfrag} shows a dense section of
the fragment train, with the optically-identified fragments 
labeled on the optical image and the infrared-identified fragments
labeled on the infrared image.
Both attempts at matching fragments are indicated.
As noted above, the ephemeris positions do not often correspond with
fragments in the {\it Spitzer} image. In Fig.~\ref{birtfrag},
fragment 73P-H matches very well. The ephemeris positions
of M and N are oriented similarly to two nearby fragments in the 
{\it Spitzer} image, and AN is close to a pair of fragments;
however, none of these correspondences are convincing.
The ordering and labeling by P. Birtwhistle allows possible
identifications for M, BO, N, H, BQ, BP, and AN with {\it Spitzer}
fragments S24, S26, S27, S28, S29, S31, and S32, respectively.
Only the correspondence for fragment H agrees with that determined
from the Horizons ephemeris, so we consider this one fragment
to have a reasonably certain cross-identification with S28.
The other suggestive identifications are indicated with `??'
in Table~\ref{fragtab}.

The lack of agreement between the ephemeris-predicted and 
{\it Spitzer}-observed positions is surprising, since the optical astrometry of small fragments that was used in the arc for the orbit solution stretches typically from 2006 Mar 24 through 2006 Apr 20, similar to the dates of the {\it Spitzer} observations. The estimated $3\sigma$ error ellipse for the predicted positions varies depending on the length of the data arc; for example for fragments
R, N, M, AB, BR, and Q,  
it is $28\arcsec\times 0.2\arcsec$, $30\arcsec\times 0.3\arcsec$, $64\arcsec\times 3\arcsec$, $110\arcsec\times 0.7\arcsec$, $330\arcsec\times 0.8\arcsec$, and
$1517\arcsec\times 1.5\arcsec$, respectively. Even taking into account these large uncertainties, the
dispersion perpendicular to the orbit is significantly larger for the predicted ($600''$) than observed ($36''$) positions,
and there should be more fragments with matches within $100''$ than observed.

Why do the predicted positions not match the infrared fragments? Some possibilities include the following:
(1) The ephemeris calculations are in error. This could arise due to the low-quality input data (inaccurate positions, which seem unlikely given the quality of
the input data and observers involved), 
or underestimated uncertainty when observations over a short arc from Earth are used to predict the locations of an object as viewed from a distance platform (given that Spitzer was 0.32 AU from Earth in May 2006).
(2) Fragment identities were incorrectly matched to astrometry, e.g. different fragments were considered together as same fragment, or one fragment was artificially split into multiple short (uncertain) arcs.
(3) Non-gravitational forces are significantly different from those of typical comets, with much higher perturbation perpendicular to the orbital plane 
($A_{3}$) 
as well as within the plane.
(4) The fragments do not last more than $\sim 2$ weeks, or are so highly variable in their
outgassing that they are only briefly detectable, so that a different set of fragments was seen at each epoch.
All of these explanations seem plausible for some fragments and may contribute to the lack of correspondence. 
We consider the last one in more detail here, since it leads to an important conclusion for the fragment longevity and evolution.

\def\extra{
frag	3sig	PA
G		5x.12	-38
R		28x0.2	-40	2006-Mar-24 to 2006-May-02
M		64x.3	-40	2006-Mar-23 to 2006-May-06
N		30x0.3			 2006-Mar-23 to 2006-May-03
AT		503x1		2006-Apr-07 to 2006-Apr-28
BR		330x.8		apr20-may1
Q		1517x1.5			2006-Mar-23 to 2006-Apr-09
E		7.3x1.8	-25		1989-Dec-09 to 2000-Dec-12
}

The longevity of the fragments, and their brightness variations, can be surmised from amateur photometry compiled by the Minor Planet Center. The primary fragment C has a typical brightening and fading as a function of heliocentric and geocentric distances. Fragment B had a significant increase around 2006 Apr 1, followed by a decrease back to approximately the same trend that existed prior to the outburst. Fragment G had a dramatic increase in brightness around 2006 Apr 8, then faded rapidly---much more rapidly than a typical trend that would have peaked near close approach and stayed bright at perihelion, so that we suspect G  has completely disrupted. 
Indeed, the HST images show G to be no more than a set of intermediate-sized fragments on 2006 Apr 19 \citep{weaverSW3}.
The miscellaneous other fragments peaked in apparent brightness around close approach to Earth and then they too faded rapidly and were not seen after May 8.

Looking back to the 2001 apparition, three components were reported: 
C and B are the bright components from 1995 and again in 2006. But fragment E was for a short time brighter than B and then rapidly faded from view, decreasing in brightness after its sudden outburst while B and C continued their gradual increase toward perihelion. We suspect fragment E experienced a catastrophic disruption in October 2001 that made it briefly 
very bright (as an unresolved swarm of icy fragments) then rapidly fade from view.

Thus it is plausible that `lettered' fragments varied significantly in sublimation rate and sub-fragmentation, making their identifications problematic and their non-gravitational forces unpredictable. The orbits of the lettered fragments (other than C and B) are highly uncertain,
and the infrared-detected fragment population (55 fragments observed together in one
mosaic) differs significantly in distribution relative to the fragment C and B orbital planes.

\def\extra{
The intermediate-sized fragments (as defined below) emanating from B were observed by HST over 3 days, during which time the individual objects' brightnesses varied erratically: examples of individual objects on days 1/2/3 include nearly all combinations of relative brightnesses (bright/bright/bright, bright/bright/faint, bright/faint/bright, bright/faint/faint, etc).
}

\clearpage

\section{Properties of meteoroids}

The surface brightness of the trail on 2006 Apr 1 was measured from the 
IRAC 8 $\mu$m and MIPS 24 $\mu$m images. 
At 8 (24) $\mu$m, the trail  surface brightness in a region containing no 
significant fragments is $0.055\pm 0.007$
($0.18\pm 0.015$) MJy~sr$^{-1}$,
the width on the sky is 148 (156)$\arcsec$, and
the width correcting for distance is 
6.9 (6.8) $\times 10^{4}$ km.
Applying extended emission 
\citep[$\times 0.737$,][]{reachCal}
and color ($/1.06$) corrections to the IRAC brightness;
a color ($/0.968$) correction to the MIPS brightness;
and scaling the IRAC brightness for the relative heliocentric
distance difference to the MIPS observation (3 days later), 
the observed
brightness ratio corresponds to a color temperature between
the nominal wavelengths of 7.872 and 23.68 $\mu$m of 
$T=261\pm 8$ K. Scaling for convenience to 1 AU, assuming 
$T\propto r^{0.5}$ yields $T_{1}=301\pm 8$ K.
The trail is not detected at 70 $\mu$m in the 2006 Apr 1 image; the upper limit at
70 $\mu$m is 0.07 MJy~sr$^{-1}$. Combining with the 24 $\mu$m brightness, we can
place a lower limit to the temperature of $>225$ K,
assuming greybody grains. Scaling for the heliocentric distance,
we obtain a lower limit to the effective temperature at 1 AU of
$T_{1}>259$ K. This limit is consistent with the temperature inferred from the
8$\mu$m/24$\mu$m ratio.

\def\extra{
from jeremie:
Warning: the width of the stream in May suggest it is twice as large: 9.3 arcmin =~ 1.5 e+5 km. Not a big deal in itself but it just shows that the observations do not show everything. As a comparison to these numbers, the tau-Herculids activity lasts for a full month, spanning over 78e+6 km of the Earth orbit (of course the inclination of the meteoroids has a role in here, but still...). }

The debris trail is warmer than expected for
dark, isothermal particles at 1 AU ($T_{1,iso}=278$ K).
The same result was found for comet trails observed at 12, 25, and 60 $\mu$m with {\it IRAS} by \citet{sykesbook}, who showed that the elevated temperature is consistent with that expected for rapidly-rotating, randomly-oriented spheres of low thermal conductivity (so they maintain a latitudinal temperature gradient).

In principle, the elevated temperature could be due to physical optics, with the grains emitting less efficiently in the infrared than they absorb visible light; however, this possibility is definitively ruled out by the particle dynamics: particles small enough ($<20 \mu$m radius) for physical optics to be important feel such strong radiation pressure that they cannot reside in the debris trail and instead are swept back and form the dust {\it tail} 
(with radiation pressure comparable to gravity).
The distribution of meteoroids is diagnostic of their size and emission history. A detailed analysis with comparison to theoretical models is presented in a companion paper 
\citep{vaubaillSW3}. In summary, the meteoroids in the debris trail are of order cm in size. The size distribution slope is similar for the meteoroids and
in the size distribution around 100 $\mu$m to 1 mm such that larger
meteoroids are much more abundant than would be guessed by taking
the coma particle size distribution and extrapolating up to meteoroid sizes.

\begin{figure}
\epsscale{1}
\plotone{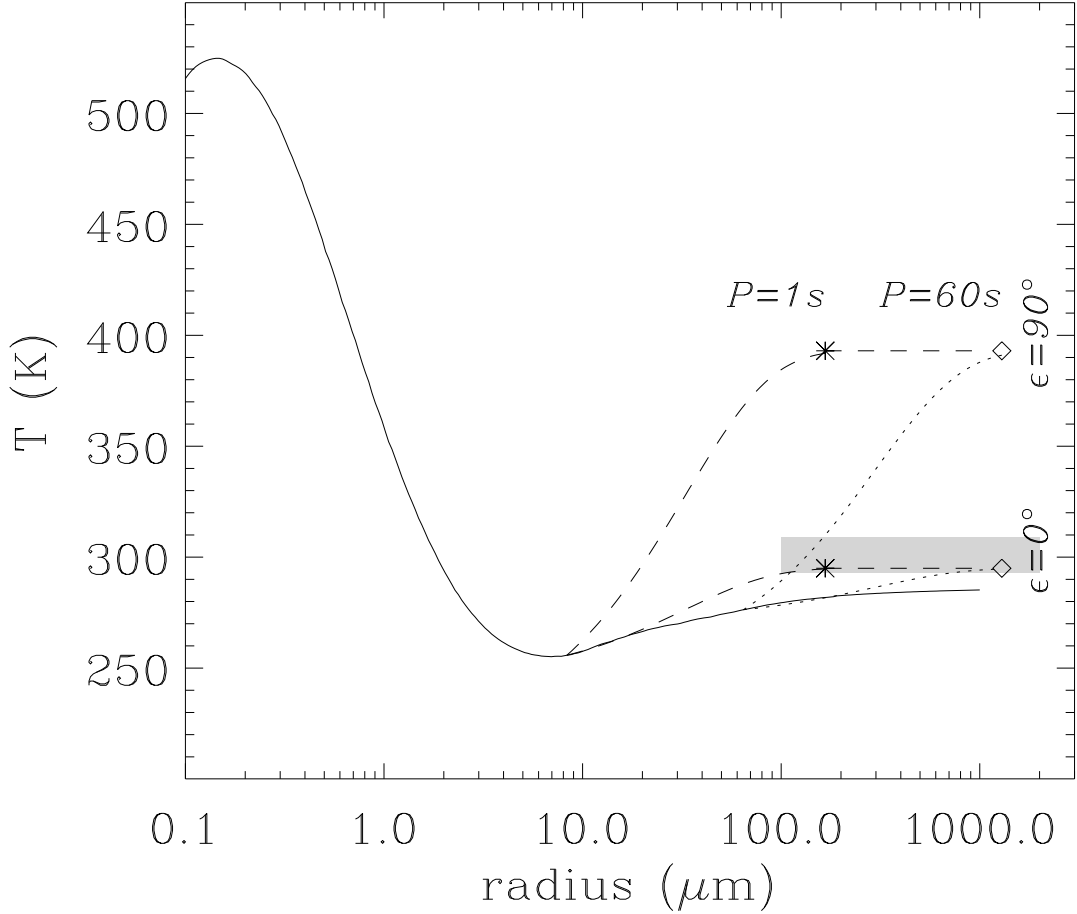}
\epsscale{1}
\caption{Temperature versus size for spherical particles illuminated by the Sun at 1 AU. 
The solid curve was calculated using Mie
theory for amorphous olivine. The other curves illustrate the behavior expected for 
particles with different surface temperature distributions. The asterisks are located
at the skin depth appropriate for particle rotating once per second.
The upper asterisk
is for particles with obliquity ($\epsilon$) of $90^{\circ}$, i.e. spin axis in the orbital plane;
the lower asterisk is for $\epsilon=0$, i.e.  spin axis perpendicular to the orbital plane.
The diamonds are located at the skin depth appropriate for particles rotating once per minute.
Dashed and dotted lines are simple interpolations between the isothermal limit (for
amorphous olivine) and the temperatures of particles with 
skin depths for 1 sec and 1 min periods, respectively.
The $2\sigma$ confidence interval for the observed temperature is shaded as a grey rectangle, 
truncated at 100 $\mu$m particle size based on dynamical constraints.
\label{tempsize}}
\end{figure}

The elevated infrared color
temperature is more likely due to the particles being much
larger than the thermal
skin depth, such that a temperature gradient is maintained across the surface.
For spinning bodies illuminated by sunlight, only those small
enough for heat to be distributed
from the day to night side more rapidly than the rotation period
will be isothermal.
For solid, rocky bodies the thermal skin depth is approximately
$L_{s}=1.7\times 10^{3} P^{0.5}$ $\mu$m where $P$ is the 
spin period in seconds.  A particle with diameter much smaller than
the skin depth will be isothermal and can be treated as a dust grain,
while a particle much larger than the skin depth should be treated
like an asteroid. Particles with size comparable to the skin
depth will have intermediate temperature distributions. 
To illustrate and bound the effects, we calculated temperatures of
spherical, isothermal grains of amorphous olivine 
\citep[optical properties from][]{dorschner95} and compared
them to the observed temperature and those expected for 
particles with temperature gradients.
Figure~\ref{tempsize} shows predicted temperature versus size.
In the large-particle (asteroid) limit, the surface temperatures
range from the subsolar temperature for a high-obliquity
body (spin axis in the orbital plane; equivalent
to no rotation or zero conductivity) to the isothermal-latitude
model applicable to a very fast-rotating body 
\citep[cf.][]{neatm}. 
Since dynamical arguments rule out micron-sized grains,
the models that can explain the observed temperature are those 
with low spin rates or moderate-to-low obliquity. A random
distribution of spin axes,
as inferred by \citet{sykesbook}, 
is consistent with the observed temperature if the spin
rate of mm-sized particles is less than 1 min.
(Slower spins would yield higher temperatures than observed.
The spin states of meteoroids are not well known.
\citet{olssonsteel87} calculated the paddle-wheel torque 
on particles from solar photons, finding a period of order
1 sec for mm-sized particles exposed to solar radiation for
1 month. The spin period is predicted to
increase as the particle size squared, but it remains below 
1 min even for cm-sized particles.
The predicted spin period for particles of mm to cm size,
the size range inferred from the dynamics 
modeling to contribute most of the debris
trail emission \citep{vaubaillSW3}, is consistent with the
observed trail temperature.

\clearpage

\section{Dynamics and survival of fragments and meteoroids\label{dynamicssection}}

We now address the dynamics of the fragments. We differentiate between the primary  (B and C) fragments, the {\it large fragments}
(lettered by the Minor Planet Center and visible as a distinct miniature comet  in the Spitzer images), the 
{\it small fragments} (visible only as individual entities in the Hubble images and otherwise as the lumpy inner coma of ground-based optical images), and the {\it meteoroids} or debris stream (visible as extended emission along the orbit only in the Spitzer image).

\subsection{Sublimation and survival}

The fragments are undoubtedly composed of ice and refractory material. A fraction of the  surface
$f_{ice}$ is exposed, and  the  sublimation rate of the ice is defined
\begin{equation}
Z \equiv \frac{\mu^{-1}dM}{dAdt}\,\,{\rm mol~cm}^{-2}~{\rm s}^{-1}
\end{equation}
where $dM$ is the mass of ice produced per unit surface area $dA$ and time $dt$, and
$\mu$ is the molecular weight of the ice
(presumably H$_2$O).
The time that a fragment can survive before its ice is sublimated is
\begin{equation}
t_s = \frac{M}{\mu Z A} = \frac{4 \rho R}{3\mu Z f_{ice}}
\end{equation}
which in physical units becomes
\begin{equation}
t_s = 54 f_{ice}^{-1} \left(\frac{\rho}{{\rm g~cm}^{-3}}\right)
\left(\frac{R}{\rm cm}\right)
\left(\frac{18 m_H}{\mu}\right)
Z_{16}^{-1} \,\,{\rm days},\label{tsurvival}
\end{equation}
where $Z_{16}$ is the sublimation rate in 
$10^{16}$ cm$^{-2}$ s$^{-2}$, $R$ is the fragment radius, 
and $\rho$ is the fragment mass density.
The sublimation rate depends on the particle temperature. A surface of pure water ice illuminated at 1 AU from the Sun will have 
$Z_{16}=1.7\times 10^{2}$ \citep{delsemme}.
The sublimation rate at the surface of a comet depends on 
its temperature distribution, spin rate, conductivity,
and composition, with estimated values of order
$Z_{16}\sim 10$--30 \citep{whipplehuebner}.
Small ice particles are colder than a large slab and hence
sublimate more slowly, with $Z_{16}\sim 0.2$ for $R=0.1$ cm
and  $Z_{16}\sim 7\times 10^{-4}$ for $R=0.01$ cm
\citep{patashnick}. 
Heterogeneous particles may however be warmer, with
$Z_{16}\sim 10^{3}$ for $R=0.5$ cm \citep{mukai}. 

The small fragments must survive at least over the 2 HST orbits during which they were observed to move rapidly but could be traced as distinct individuals; setting $t_s>2$ orbits requires $R>400 f_{ice}$ cm. This condition is likely to apply since the inferred size of the small HST fragments is much larger; however, the HST magnitudes sizes could be overestimated if they are dominated by comae. The HST fragments that disappeared between observations (or changed into diffuse clouds) probably disaggregated into debris, visible in the images due to dust scattering sunlight, rather than sublimating into vapor. 

The debris trail particles must survive years in order to occupy their observed locations along the projected orbit; this requires that the debris trail particles have radii $R> 0.6 f_{ice}Z_{16}$cm. In terms of
lifetime, small amounts of ice could be present; however, the dynamics arguments below show that $f_{ice}$ must be small in the trail particles.
\def\extra{THIS PART USED TOO HIGH SUBLIMATION RATE
Only particles smaller than 10 cm can be levitated from the comet's surface by gas drag (see next paragraph). Thus in order to survive as long as they do, the debris trail particles must have very small $f_{ice}$: either they started with a small amount of ice that was sublimated without disaggregating the particle, or they formed from relatively `ice-free' portions of the comet's mantle.
}

The basic model for cometary grain production has the  expanding gas from sublimating ice near the surface carry off solid materials that are entrained 
if the drag force is greater than the gravitational pull of the nucleus:
\begin{equation}
\frac{1}{2}
\mu v Z  \pi a^2 > \frac{G M m}{R_N^2}
\end{equation}
where $v$ is the  expansion velocity of the vapors from the sublimating ice,
$a$ and $m$ are the particle radius and mass, 
$R_N$ and $M$ are the nucleus radius and mass, 
and $G$ is the gravitational force constant. 
The maximum size that can be levitated
is then 
\begin{equation}
a_{max} =  \frac{9 \mu v Z}{32 \pi G \rho_N \rho R_N}
\end{equation}
where $a$ and $\rho$ are the grain radius and density, and $\rho_N$ is the nucleus density. In physical units, for a nucleus
with sublimation rate $Z_{16}=20$,
\begin{equation}
a_{max} \sim 10 
\left(\frac{R_N}{\rm  km}\right)^{-1}\,\,{\rm cm},
\end{equation}
for $\rho=\rho_{N}=1$ g~cm$^{-3}$.
Since 73P has perihelion $q=0.94$ AU and the nuclear radius in 1994, before the recent breakup events, is $R_N<1.1$ km \citep{boehnhardt99}, 
the largest particles that can be levitated by gas drag are
of order  10 cm in size.

After decoupling from the gas, the particle's trajectory is determined by the velocity that has been imparted to it by gas drag, $v_{ej}$, and the ratio of radiation pressure force to solar gravity, 
\begin{equation}
\beta\equiv F_{rad}/F_{grav} =
0.57 Q_{pr}\rho^{-1}(1\,\mu{\rm m}/a)
\end{equation}
for spherical particles,
where $Q_{pr}$ is the a coupling efficiency
between the particle and solar radiation \citep{burnslamysoter};
for particles much larger than the wavelength of visible light,
i.e. $a\gg 1$ $\mu$m, we can set $Q_{pr}=1$.
For fragmentation or disaggregation of a nucleus (or meteoroid), the treatment is analogous to that of debris, with the velocity imparted to the fragment (due to whatever process caused the fragmentation) taken to be $v_{ej}$.

For small objects, and additional term must be added to account for the `rocket' effect due to expansion of vapors from sublimation of ice on its surface. To explore this effect, we define
\begin{equation}
\alpha\equiv \frac{F_{rocket}}{F_{grav}} = \frac{3 \mu Z v_{ice} f_{ice}}
{4 G M_\odot\rho a}
\end{equation}
where $v_{ice}$ is the velocity of expansion of the sublimating ice. 
In physical units,
\begin{equation}
\alpha = 0.036 \left(\frac{v_{ice}}{{\rm km~s}^{-1}}\right)
\left(\frac{{\rm g~cm}^{-3}}{\rho}\right) 
Z_{16} 
\left(\frac{1 {\rm ~cm}}{a}\right) 
f_{ice}. \label{eq:alpha}
\end{equation}
Using this parameterization, the rocket effect $\alpha$ is analogous to 
radiation pressure $\beta$: both represent forces that point away from the Sun and
are fractions of solar gravity.
It is furthermore useful to make the ejection velocity dimensionless
through $\nu\equiv v_{ej}/v_{orb}$, where $v_{orb}$ is the orbital speed at the time of particle production (normally, close to perihelion).
The parameters $\beta$, $\alpha$, $\nu$ allow a dimensionless means to
determine the dominant celestial mechanical perturbation for a particle.
(Note that unlike the radiation pressure parameter, $\beta$, the rocket parameter,
$\alpha$, depends on heliocentric distance. This is due to the fact that
while insolation and radiation pressure decrease with distance from
the Sun as $r^{-2}$, the sublimation rate has a steeper dependence taking
into account the need for the ice to be heated above its latent heat before
it will sublimate. Furthermore note that we assume the outgassing is in the direction
of the Sun, so that this parameterization neglects the the rocket force due to the component 
of the outgassing that is nonradial.)

\begin{figure}
\epsscale{0.34}
\plotone{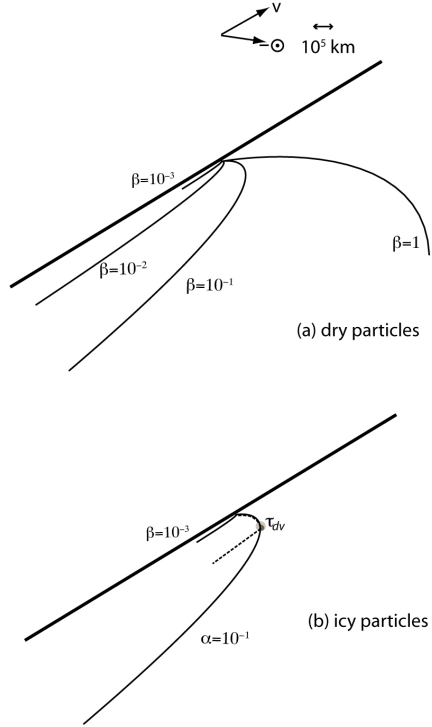}
\epsscale{1}
\caption{Illustration of dynamics for (a) dry particles and (b) icy particles.
The arrows at the top indicate the direction of the velocity vector ($v$)
and the antisolar direction ($-\odot$) projected onto the observed sky plane.
The dynamics of dry particles is determined only by the ratio of radiation pressure to solar gravity, $\beta$. Particles in the anti-solar direction, such as observed in the tails of the 73P-B, 73P-C, etc fragments, have $\beta>10^{-1}$
corresponding to sizes $< 10$ $\mu$m. Particles along the debris trail
(which closely follows the projected orbit) must have $\beta<10^{-4}$,
corresponding to sizes $>1$ cm.
The dynamics of icy particles is determined by the ratio of rocket force 
(from sublimating ices) to
solar gravity, $\alpha$, as well as $\beta$. The particle illustrated in
panel (b) has a radius 1 mm, $\beta=10^{-3}$ and begins with 
$f_{ice}=0.0017$ so $\alpha=10^{-1}$.
Since initially $\alpha\gg\beta>\nu$, the particle follows the 
syndne for $\alpha=10^{-1}$.
The particle becomes devolatilized 
(no more ice exposed to sunlight) at time $\tau_{dv}$, 
so $\alpha\rightarrow 0$,
and the remaining,
effectively dry particle then follows a $\beta=10^{-3}$ trajectory.
Debris trail particles follow the projected orbit closely and therefore
must have $\tau_{dv}\rightarrow 0$. The small fragments in the HST images
follow the tail-ward (large $\alpha$) direction, then either disintegrate
or follow low-$\beta$ trajectories so far from the parent nucleus that
they are effectively lost from the system.
\label{alphacartoon}}
\end{figure}

\clearpage

\subsection{Dynamical classes of fragments}

The trajectories determined by the combination of $\nu$ and
$\alpha+\beta$ can be divided into classes. 
{\it Large fragments} have 
$\alpha+\beta\simeq 0$; they remain close to the projected orbit,
with trajectories determined only by $\nu$. (Gravitational perturbations by the planets are of course important  on long timescales. Here we consider only the 3 orbits during which the fragmentation events transpired and during which no close encounters with planets occurred.) 
The separation velocities for fragments of split comets have been deduced in the range of 1-2 m~s$^{-1}$ \citep{sekanina78}, 
corresponding to $\nu\sim 10^{-5}$. 
For comparison, the radiation pressure has $\beta<10^{-5}$ for
particles with radii $a> 6\rho^{-1}$ cm.
{\it Intermediate-sized, icy fragments} appear in clusters around large fragments (perhaps being their principal building blocks) and have dynamics, relative to their local center of mass, dominated by 
$\alpha$, the rocket effect; the size scale is of order 10 m. 
These fragments are discussed in more detail in a paper describing
the HST results; the dynamical modeling and discussion are included
in this paper in order to contrast these fragments from the others.
{\it Debris trail particles} are smaller and lie along the orbit 
of the parent comet; they have evidently $\alpha\ll \beta$ so that
their dynamics is dominated by radiation pressure.
In the following subsections we examine the dynamics of each type of 
fragment in more detail.

\subsection{Separation of large fragments}

The small perturbation caused by the radial and in-orbit 
components, $v_{\parallel}$, of the ejection velocity
causes large fragments with $\alpha + \beta \simeq 0$ to spread along the orbit of the nucleus,
while the perpendicular component,
$v_{\perp}$,  causes fragments to disperse perpendicular to the orbit.
We can therefore
use the observed locations of the fragments perpendicular to the orbit to estimate $v_{\parallel}$
and $v_{\perp}$.
A natural but wrong way to compute these quantities is to simply divide the distance between
the fragment and the comet by the time since its ejection;
however, care must be taken here.
A fragment ejected with the velocity $v_{ej}$
at a point $P$ of the cometary orbit will have a slightly different inclination $i$ and semi-major axis $a$
than the main body.
Considering Keplerian motion, this implies that the fragment will always pass by point $P$
at each subsequent passage. As a consequence, the physical separation
between the fragment and the main body will vary with the true anomaly.
We assume all fragments were produced in the 1995 breakup event, and
we determine $v_{\parallel}$ and $v_{\perp}$ by matching the location
 of the fragments by iteration.
Because $v_{\parallel}$ affects $a$, it has a strong influence on the location of the fragments within the orbit plane.
Nearly all the large fragments are located closer to the cometary orbital plane,
and their inferred $v_{\parallel}\sim 1$ m~s$^{-1}$. 
Interestingly, their inferred $v_{\perp}$ 
is similar to $v_{\parallel}$, suggesting 
the ejection from the parent comet was
roughly isotropic.

One fragment, illustrated in the inset at the top of Fig. 3b, 
is located far from the comet orbital plane.
Its distance from fragment B in the plane is 
$1.9^{\circ}$ (or $1.8\times 10^6$ km),
and it is $4.1'$ (or $6.6 \times 10^4$ km) away from the orbit plane.
Assuming that this fragment was ejected during the 1995 breakup event,
we find $v_{\parallel} \sim 1.4$ m~s$^{-1}$ and 
$v_{\perp} \sim 16$ m~s$^{-1}$.
The value of $v_{\perp}$ is very high, relative to 
$v_{\parallel}$ for this fragment and 
inferred ejection velocities of other
comet fragments.
(All the other fragments are located much closer to the cometary orbital plane,
with corresponding $v_{\perp}$ is less than 
2 m~s$^{-1}$.)

The location of a fragment can be influenced by non-gravitational
force from asymmetric outgassing ($\alpha \neq 0$).
Translated into the Marsden (1961) 
non-gravitational force parameter, we find 
$A_{3}\simlt 5.0 \times 10^{-9}$ for all fragments except the one
discussed in the previous paragraph. 
This value is higher than typical Jupiter-family comets,
where the median absolute value is 
$|A_{3}|=1\times 10^{-9}$.
The fragment located furthest from the orbital plane has
$A_2=8.9 \times 10^{-9}$ and $A_3=-2.0 \; 10^{-7}$. 
The later value is one order of magnitude
higher than the highest value recorded for cometary nuclei, namely comet C/1996 N1 Brewington
($A_3=-1.3 \; 10^{-8}$ ; P. Rocher, IMCCE, cometary note 0209\footnote{http://www.imcce.fr/en/ephemerides/donnees/comets/FICH/CIA0209.php}).
however given the size of the fragment one expects that the outgassing process will have a strong influence on its dynamics as discussed above for the 
mini-fragments in the HST images.

The reality is that a combination of ejection velocity and 
non-gravitational forces determine the fragment locations.
Our findings show that the fragments observed by {\it Spitzer} have high but reasonable ejection speeds. One of the fragments must have been released at very
high speed perpendicular to the orbit. 
It must be stressed that none of these fragments could have been ejected 
in 2006---otherwise, the ejection velocity or $A_i$ values would be far 
higher than the values reported here.

\subsection{Rocket effect and HST fragments}

The rocket effect is more important than initial ejection velocity
($\alpha>\nu$) for particles with
$a\rho<3.6 f_{ice}v_{ej}$ km.
That  non-gravitational rocket forces are significant for objects the size 
of cometary nuclei is well-known, but the ice-sorting effect on
fragments is a new application.
The effect is analogous to radiation pressure in direction, but is approximately 
$600 Z_{16} f_{ice}$ times larger, so the rocket effect and radiation pressure are equivalent for $f_{ice}=10^{-2}$. 
The small fragments seen in the HST images are
estimated to have sizes of order 5 m.
Some of them are visible
in 3 consecutive HST orbits, so they survive at least 
at least 200 min. Their apparent motion on the sky
is $\sim 100$ m~s$^{-1}$ in the antisolar direction---the same direction
as the dust and gas tails  \citep{weaverSW3}.
The direction of motion clearly indicates $\alpha+\beta>10^{-2}$, and
the size indicates $\beta\ll \alpha$. Combining the size and $\alpha$,
we solve for the fraction of the surface of the small fragments
that must be covered by ice
\begin{equation}
f_{ice}>0.8 \left(\frac{a}{5 {\rm~m}}\right) 
\left(\frac{{\rm km~s}^{-1}}{v_{ice}}\right),
\end{equation}
so the small fragments must be very icy in order to move in the tailward direction. It is possible that the sizes of the fragments are overestimated
from the HST images due to their bright comae, but the presence of those
comae demonstrate that the fragments are very active and icy;
thus $f_{ice}$ is unlikely to be smaller than 0.1.

A significant difference between the rocket and radiation pressure effects is
that the former may be time-dependent, if the fraction of exposed ice
changes with time. We discuss this in more detail in
the next subsection with regard to the meteoroid-sized particles.
The fate of the icy fragments is to either (1) continue along high-$\alpha$
trajectories, (2) become `extinct' and transfer to orbits roughly similar to the
parent comet, or (3) disaggregate into clouds of dust and gas. 
Even in cases (1) and (2), it is clear that the small fragments 
are {\it lost} from the cometary system, in the sense that their orbits are
so different (and the objects are so small) that they would be unlikely to be detected again and even if so their orbits have already been sufficiently perturbed from that of the parent comet that they will be many degrees from that comet upon return and would not be noticed in
observations centered on the comet. Thus a
general result is that fragments in the $\sim 5$ m range are a form of missing
mass that could only have been detected thanks to the high angular 
resolution of HST and the close approach of 73P to the Earth.
Observations of another fragmenting comet LINEAR indicated that the
detectable fragments and dust do not add up to the comet's pre-breakup
mass \citep{weaverlinear}. It now appears that a significant portion of
the cometary fragmentation may be in the form of icy fragments
in the 5-m size range.

\def\extra{\tt\bf ??? insert some more precise numbers???}

Another aspect of the small fragments seen with HST is that some of them
disappear in between successive images, and their are some fuzzy patches
without nuclei that move away from the comet. 
This fragmentation and disaggregation appears to be analogous to that of the main fragments, except that for the smaller ones the there
remains only an a cloud of vapor. The material strength  of the fragments is therefore probably small. 
Based on the tidal splitting of Shoemaker-Levy 9, the
strength of the nucleus was estimated to be $\sim 10^4$ dyne~cm$^{-2}$
\citep{asphaug}.
If the small fragments have strengths of this order, then they are
even more likely to fragment and disintegrate because they have
far less gravity than the nucleus and cannot support themselves 
as `rubble piles' once they have been damaged by collision or thermal stress.

\def\extra{\tt\bf These HST fragments are a source of extended gas production....}
 
\subsection{Radiation pressure and debris trail}

The debris trail in Figures~\ref{mar06min} and ~\ref{monmos} clearly shows the presence of long-lived debris within the comet's orbit. We can constrain $\alpha+\beta<10^{-5}$
from  the location and  width of the trail \citep{vaubaillSW3}. If the particles
were levitated from the nucleus, as in normal cometary activity \citep{reachtrail}, then they have radii $<10$ cm and $\beta>6\times 10^{-6}$. 
Thus only the largest particles that can be levitated from the nucleus
reside within the debris trail, and
there is no room for these particles to have significant rocket effect.

Larger particles could be present in the debris trail if they were 
released from the nucleus by another mechanism like nuclear fragmentation. 
They must still achieve escape velocity from the nucleus, 
$v_{esc}\simeq 1 (R_{N}/{\rm km})$ m~s$^{-1}$, which corresponds to a velocity
perturbation parameter $\nu> 2\times 10^{-7}$. Meteoroids of order 1 m size could
be present in the debris trail with small amounts of ice as long as 
$\alpha<10^{-5}$. Referring to Eq.~\ref{eq:alpha}, we can then constrain 
the sublimation rate $Z_{16}f_{ice}<0.03$. The size range discussed in
this paragraph defines a transition between the icy fragments that follow
the dust tail, like those seen in the HST images of 73P, and smaller
and lower-sublimation-rate bodies. Note that the observed surface brightness
of debris trails is unlikely to be produced by m-sized chunks; the required
mass of debris would be too high. Furthermore, meteor showers are due to 
$\sim$mm-sized meteoroids. The m-sized fragments would be rare denizens of the trails.
 
As volatiles disappear from the
surface of a debris trail particle, its interior may eventually become
either dry or shielded by refractory material remaining at the edge of the particle. 
The situation is somewhat analogous to that of the comet as a whole, 
where a `rubble mantle' of particles that cannot be levitated eventually shields the  interior from sunlight and reduces the active area so $f_{ice}\ll 1$ and the activity and its 
resulting non-gravitational force terminate. 
Cometary debris particles have negligible gravity, but their material strength may be significant. 

If the  refractory part of a particle had no material strength at all, e.g. if it were an agglomeration of very small particles  held together  only by the ice, then the particles 
would disaggregate while they are sublimating. The tiny subunits would flow away at the same time as the gas, and the particle would simply shrink until only one subunit is left. 
If one believes this
model for the cometary debris constitution, then $\alpha$ is always large, it
dominates radiation pressure, and the lifetime is set by 
Equation~\ref{tsurvival}. 
Cometary debris trail particles survive for $>10$ yr
\citep[two orbital revolutions][]{sykeswalker,reachtrail}.
Averaging over a typical short-period comet orbit with 5 yr period, 
$\langle Z_{16}\rangle\simeq 0.2 Z_{16,1}$,
where $Z_{16,1}$ is the sublimation rate at 1 AU.
For particles with $Z_{16,1}\sim 1$ to survive for $>10$ yr, they
must be larger than 14 cm in size. Such particles could not have been
levitated from the nucleus by gas drag,
and the mass of such particles required to explain the observed, smooth and bright, 
debris trails would be orders
of magnitude greater than that of the nucleus from which they originate.
Meteoroids producing known
meteor showers survive for $>10^3$ yr
\citep{jenniskens}; only bodies much larger than 1 m could survive 
continuous sublimation so long.

That the dynamics of the  debris trails and the small fragments are distinctly
different indicates that  their internal structures are different. The debris trail particles must be {\it dry}. Based only on the location of the particles, the size $a<10$ cm, and 
$\alpha<10^{-5}$, so $f_{ice}<2\times 10^{-5}$. 
The survival of the debris trail particles requires that they either originated from a completely dry portion of the nucleus or they have become devolatilized almost immediately after separation from the nucleus. 
If the particles originated from a completely dry portion of the comet, then we must
explain {\it how} they were ejected from the comet.
Possibly the particles were from the mantle, and they
were lifted from the surface upon activation of a patch of buried ice. 
Alternatively, the particles could be from a part of the nucleus adjacent to an
active region, and they fell into or were otherwise entrained in the outflow.
Both of these scenarios and the subsequent dynamics of the particles are consistent with
the interplanetary dust particles of cometary origin having
anhydrous mineralogy \citep{messenger06} and the surfaces
of 9P/Tempel 1 and 81P/Wild 2 appearing predominantly `dry' in
recent spacecraft fly-bys \citep{ahearnT1,brownlee04}.
If the particles did maintain some volatile compounds, then
in the process of devolatilization upon release from the surface and exposure to the full ultraviolet radiation of the Sun, the refractory parts of the particle must have retained strength. That is, they could not be composed of 
very fine ($<10$ $\mu$m) grains that are held together only by volatiles
acting as a sort of `glue.' 
Thus the debris trail particles (and meteoroids of the size
that produce meteors) must have  some material strength.

This inference, based only on dynamics, is compatible with knowledge of
the properties of meteors and interplanetary dust particles. 
Of some importance for understanding the nature of comets and interpreting
collected, or remotely observed, small ($<100$ $\mu$m radius) interplanetary particles in comparison to the larger ($>1$ mm) particles that dominate the
total mass loss, is whether the larger particles are amalgams of the smaller
particles, whether the small particles are amalgams of fine grains, and
how those subunits are bound together.
Meteors are
known to fragment in the Earth's upper atmosphere, which
suggests they are not very strong, but in fact the impact onto the Earth's
atmosphere is an energetic event and still allows for significant strength,
estimated in the range $10^4$ dyne~cm$^2$ with the weakest ones 
at $\sim 4\times 10^2$ dyne~cm$^{-2}$ \citep{trigo}.
The stress on the particle's refractory structure can be roughly
estimated from the sublimation rate 
$Z v \mu m_H \sim 5$ dyne~cm$^{-2}$ for sublimation at 1 AU from the Sun.
The sublimation rate will scale as $r^{-2}$, so particles with the strength of 
weak meteors would explode if they were released within 0.1 AU of the Sun.
(The existence of Geminid meteoroids, with perihelion 0.13 AU, argue for a 
high strength for bodies at $\sim$mm sizes.)
But for a short-period comet like 73P (or even the very-low-perihelion short-period comet 2P, with $q=0.339$ AU), such close approaches to the  Sun to not apply
and particles have sufficient strength to survive sublimation of their embedded
ices.

Meteor showers have been associated with comets through similarity
of their derived orbits. The meteoroid orbits are usually different
from the {\it present} orbit of the comet, and are instead associated with
past orbits. The primary acting agent in the celestial mechanics of meteoroid
streams is gravitational perturbations by Jupiter.
Even in absence of a close encounter with Jupiter, the orbit of the comet slightly changes at each return. The meteoroids are ejected on orbits similar but independent from the comet. Because of the ejection velocity and the radiation pressure, their orbital semimajor axis is typically greater than 
that of the comet. As a consequence the Jovian perturbations are not the same at each point of the stream, and the result is the deformation of the whole structure. As long as the perturbation are small enough it is still possible to associate a trail with its parent body. However, the knowledge of the passage of ejection requires a full calculation of the evolution of the stream.
The success of such calculations in predicting
the peaks of the Leonid showers confirm the basic theory \citep{leonids}.
If the meteors are ejected from the comet with some surface ice
with $f_{ice}>10^{-2}$, then their initial orbits will be affected.
Simple treatments of the physics of separation from the nucleus 
consider one parameter, the ejection velocity, which includes effects of gas drag,
decoupling, and by extension effects such as sublimation of retained ice.
The total work done on the particle is the rocket force integrated over the path until devolatilization, which we can characterize with the ratio 
\begin{equation}
w \equiv \frac{F_{rocket} \tau_{dv}} {F_{grav} P} =
\alpha \tau_{dv}/P
\end{equation}
where $P$ is the orbital period and $\tau_{dv}$ is the lifetime of ice on the surface. 
Constraints placed traditionally
upon the ejection velocity may be alternatively interpreted as a limit
on $w$ with order-of-magnitude $w\sim v_{ej}/v_{orb}=\nu$. 
A limit of $v_{ej}< 10$ m~s$^{-1}$ corresponds to
a limit of $w<0.2$ for typical orbital velocities near perihelion.
For an orbital period of 5 yr, we then find a limit $\alpha\tau_{dv}<1$ yr.
For cm-sized particles, we then find a limit $\tau_{dv} f_{ice}<0.2$ yr.
Thus if meteoroids are very icy ($f_{ice}>0.1$), they must become 
devolatilized within 2 yr. Conversely, the ice that
can remain in meteoroids that are $10^2$ yr old is $f_{ice}<2\times 10^{-3}$.


\section{Gas in the coma\label{cosection}}

\begin{figure}
\epsscale{.9}
\plotone{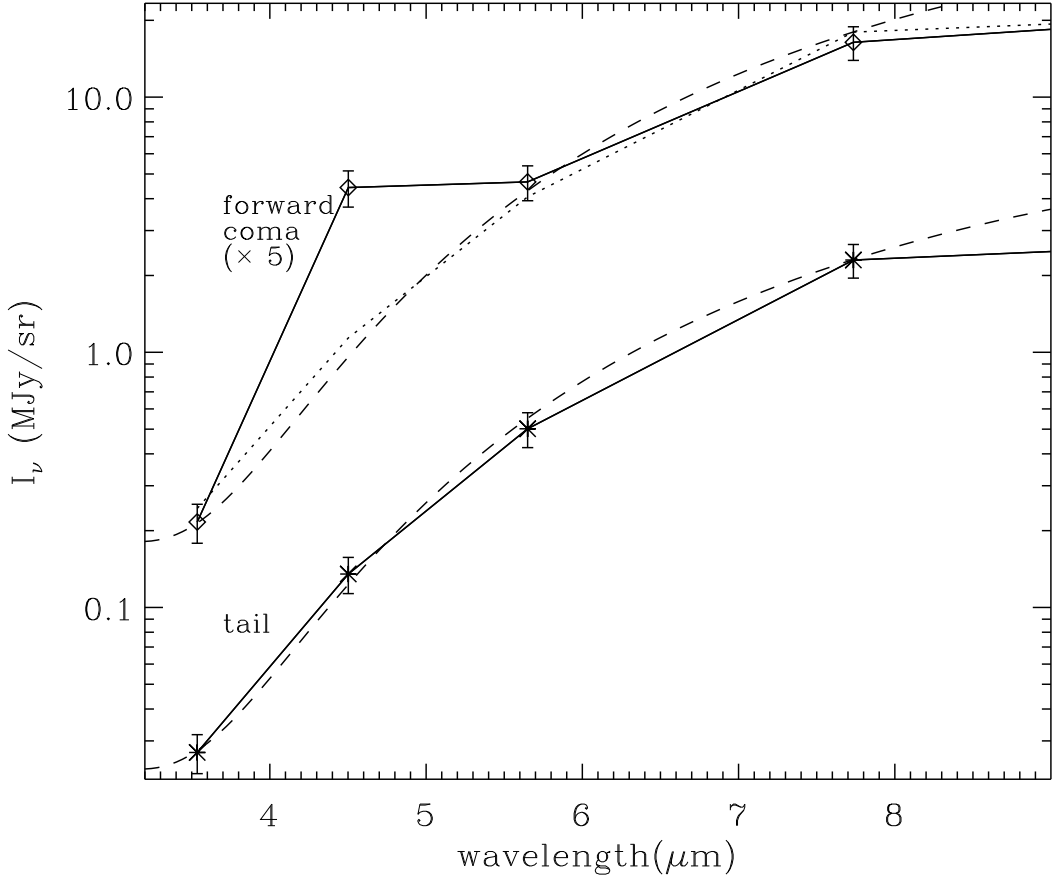}
\epsscale{1}
\caption{Spectral energy distribution from 3.6--8 $\mu$m 
obtained from IRAC photometry of two locations near 73P-B.
The upper curve is the forward coma brightness, averaged over a circular region with radius $7''$
centered $11''$ (5000 km) in the sunward direction from the nucleus; it is scaled by
a factor of 5 for clarity in this plot. 
The lower curve is the tail brightness, averaged over a 
circular region with radius $30''$ centered $197''$ (92,000 km) in the anti-sunward
direction from the nucleus. The solid line simply connects the data points.
The dashed line is a blackbody fit ($T=288$ K) to the tail brightness, shown both through the tail brightness measurements and again through 
the forward coma brightness (after scaling to match the 5.8 and 8 $\mu$m points).
The dotted line is the tail brightness, scaled upward to match the 5.8 and 8 $\mu$m
forward coma brightness.
There is a clear excess of emission in the 4.5 $\mu$m band in the forward coma,
with respect to the dust spectrum of the tail.
\label{iracsed}}
\end{figure}

The IRAC images at 4.5 $\mu$m have a morphology distinctly different from that at the other wavelengths. The spectral energy distribution (Fig.~\ref{iracsed}) clearly shows that there are locations in the coma of 73P-B with a distinct excess of 4.5 $\mu$m emission with respect to the continuum that dominates most infrared wavelengths. Figure~\ref{iracsed} shows the spectral energy distributions of a sunward portion of the coma and and anti-sunward 
portion of the dust tail. 
The continuum brightness due to dust is a combination of thermal emission and scattered light. 
The scattered light has negligible contribution except possibly at the shortest 
wavelength: 
if all the 3.6 $\mu$m brightness were scattered,
the estimated scattered light in the other bands (i.e. the 3.6 $\mu$m brightness
scaled by the solar spectrum) is $<$10\%.
To fit the thermal emission for the dust tail, we use the 5.8 and 8 $\mu$m brightness, after color
corrections are made for the wide IRAC bandpasses as appropriate for a 300 K blackbody
\citep{reachCal},
to measure a color temperature of 288 K for the tail. The scattered light
brightness required to explain the 3.6 $\mu$m emission above the 288 K blackbody
emission amounts to 50\% of the 3.6 $\mu$m brightness and 6\% of the 4.5 $\mu$m
brightness. 

For the forward coma, we remove the combined thermal emission and scattered light
assuming the spectrum of the dust is the same as in the tail. While the dust temperature is likely to vary somewhat within the coma due to particle size changes, the variations
are evidently not large, because the tail spectral energy distribution falls
very close to the coma brightness at 5.8 and 8 $\mu$m. For the region used
in Figure~\ref{iracsed}, the excess 4.5 $\mu$m brightness amounts to 
0.75 MJy~sr$^{-1}$ within the IRAC band. We attribute the excess brightness to
line emission. Based on the spectrum of 103P/Hartley 2 (a Jupiter-family comet similar 
to 73P) obtained with 
{\it ISO}, the brightest feature in all the IRAC bandpasses, by far, is the CO$_{2}$ 
$\nu_{3}$ band at 4.26 $\mu$m, with a weaker contribution from CO \citep{colangeli103P}.
The brightness of the excess emission attributed to gas features in the forward coma of 73P-B, at 5000 km from the nucleus,
is $1.1\times 10^{-4}$ erg~s$^{-1}$~cm$^{-2}$~sr$^{-1}$. 

\begin{figure}
\epsscale{.9}
\plotone{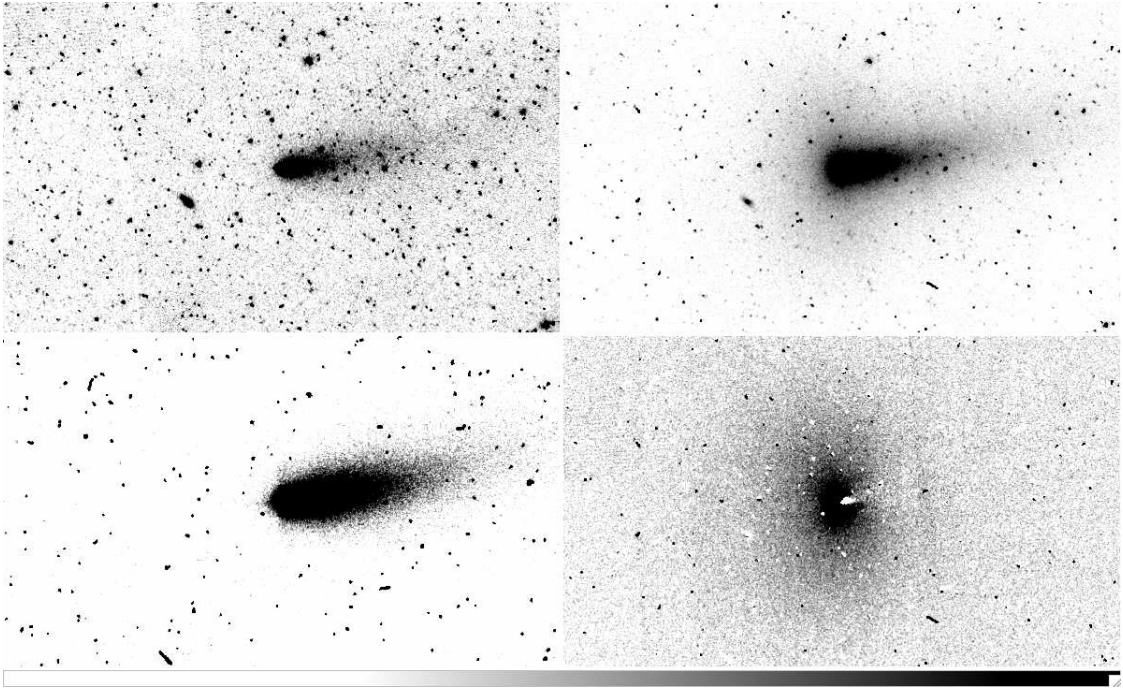}
\epsscale{1}
\caption{IRAC images of the coma and part of the dust tail of fragment B, $12'\times 7'$ in
angular size: 
{\it top left}: 3.6 $\mu$m band, 
{\it top right}: 4.5 $\mu$m band, 
{\it bottom left}: 5.8 $\mu$m band, 
{\it bottom right}: excess 4.5 $\mu$m band after
subtracting scaled 3.6 and 5.8 $\mu$m bands.
The 3.6 $\mu$m and 5.8 $\mu$m images are dominated by
dust scattering and thermal emission, respectively. The
excess 4.5 $\mu$m is due to CO and CO$_{2}$ gas emission.
\label{ch2exmap}}
\end{figure}

The distribution of line emission in the 4.5 $\mu$m band
(after removing thermal emission
using 288 K blackbody scaled to match the 5.8 and 8 $\mu$m bands and scattered sunlight using
a solar spectrum scaled to match 50\% of the 3.6 $\mu$m band) is shown in
Figure~\ref{ch2exmap}.
It is already evident from the 4.5 $\mu$m image alone that there is an extra spatial
component located in the sunward direction. The subtracted image shows the distribution 
of this excess is a very wide elliptical fan-like region primarily forward of the
nucleus. The full-width at half maximum brightness is $36''$ (17,000 km) and the
emission is detected out to $79''$ (37,000 km) in the sunward direction.

The radial profile of the CO$_{2}$ roughly follows a $\rho^{-1}$ shape, where $\rho$ is
the separation from the nucleus on the plane of the sky, as expected if the 
CO$_{2}$ molecules have a $r^{-2}$ spatial density profile and observations are
within the dissociation distance. The dissociation timescale for CO$_2$ is 5.8 days
at 1 AU from the Sun \citep{crovisier94}, so the dissociation scale length
corresponds to $6^\prime$ from the nucleus in 2006 May; that the 
dissociation length is further than the distance out to which we measured the radial profile
indicates dissociation is not expected to dominate.
A $\rho^{-1}$ profile would apply for
outflowing molecules in steady state produced from the nucleus,
as opposed to an extended source, e.g. sublimation from surface of grains or
production by photo-dissociation of larger molecules.
That the observed radial profile is consistent with a nuclear origin is
consistent with CO$_{2}$ dominating the line emission in the 
4.5 $\mu$m band, considering that CO has been found to have
an extended source in many cases, as reviewed
by \citet{bockleemorvan}.
Integrating the profile out to $48''$ radius leads to a flux 
$8.6\times 10^{-12}$ erg~cm$^{-2}$~s$^{-1}$.
The flux is a factor of 3 lower than that of the very active comet 103P
\citep{colangeli103P}, so using the same excitation scaling, 
the production rate $Q(CO_{2})\sim 1\times 10^{27}$ s$^{-1}$.
This flux must be treated with caution since the dust continuum subtraction is
inaccurate near the nucleus where the emission is brightest; the total flux also
scales linearly with the radius out to which one integrates.

For fragment C, the spectral energy distribution of the tail and coma are similar
to that of fragment B. There is also a 4.5 $\mu$m excess from the sunward side of the coma, with similar morphology, though it is lower in amplitude than that for fragment B. Applying the same
spectral and spatial fit as described above, we find a flux 
$\sim 3.0 \times 10^{-12}$ erg~cm$^{-2}$~s$^{-1}$, and an estimated 
production rate $Q(CO_{2})\sim 7\times 10^{26}$ s$^{-1}$.

The H$_{2}$O production rate of fragment B was measured by \citet{hitomi} on
2006 May 7, when the comet was at 1.03 AU from the Sun, to be $3.5\times 10^{28}$ s$^{-1}$. If both the CO$_{2}$ and H$_{2}$O production rates scale as $r^{-2}$, then
the ratio of production rates at 1 AU is
$Q_{CO_{2}}/Q_{H_{2}O}\sim 0.05$. 
The H$_{2}$O production rate of fragment C was measured by \citet{disanti07} on
2006 May 28, when the comet was at 0.95 AU from the Sun, to be $1.3\times 10^{28}$ s$^{-1}$. If both the CO$_{2}$ and H$_{2}$O production rates scale as $r^{-2}$, then
the ratio of production rates at 1 AU is
$Q_{CO_{2}}/Q_{H_{2}O}\sim 0.11$. 
Evidently CO$_{2}$ is highly abundant in the 73P-B and C ices.
The abundant CO$_{2}$ in 73P could be due to its recent exposure to sunlight since the splitting event in 1995 (and subsequent splittings for fragment B), or it could be the
typical cometary composition.
For 103P/Hartley 2, which has not split recently and was
observed at 0.84 AU from the Sun, 
the CO$_{2}$/H$_{2}$O abundance ratio is 0.08 \citep{colangeli103P},
comparable to the 73P fragments.

For comparison, the CO production rate was
found to be exceptionally low (relative to H$_2$O) 
for 73P-C: $Q_{CO}/Q_{H_{2}O}=0.005$ \citep{disanti07}.
Combining the near-infrared spectral data with our 4.5 $\mu$m result yields an
abundance ratio $Q_{CO_{2}}/Q_{CO}\sim 20$ making CO$_{2}$ the second-most-abundant
(by far) molecule in the ice exposed on 73P-C. For comet Hale-Bopp at 2.93 AU from the Sun, \citet{crovisierHaleBopp} measured production rates with relative proportions for 
H$_{2}$O:CO:CO$_{2}$ of 100:22:70. The production rate of CO$_{2}$ was very high for the
Hale-Bopp observation (even higher than we find for 73P), but that is mostly
due to the large heliocentric distance for the Hale-Bopp spectra---the H$_{2}$O
remains solid and does not sublimate rapidly at 2.9 AU. After extrapolating 
inward to 1 AU, \citet{bockleemorvan} estimated production rates for
Hale-Bopp with proportions H$_{2}$O:CO:CO$_{2}$ of 100:23:6.
Thus the CO$_{2}$ production rate relative to H$_{2}$O for 73P-C
appears to be within the range of 
other comets, and the CO is anomalously low in the 73P-C ices.

\clearpage

\section{Comparison to other comets}

\subsection{Icy fragments}
For C/1996 B2 Hyakutake, \citet{desvoivres} identified and measured trajectories of 7 distinct condensations in the coma. The separations of the fragments increase quadratically with time, clearly indicating acceleration due to non-gravitational forces and allowing an accurate measure of the acceleration and the date of splitting from the main nucleus. Taking the observed separation and time since splitting, $t$, we can determine the $\alpha$ due to the rocket effect as defined above, 
\begin{equation}
\alpha = \frac{2 d r^2}{t^2 G M_\odot},
\end{equation}
where $d$ is the separation from the nucleus corrected for the projection of the solar direction on the sky.
The 7 observed condensations have
$1.2\times 10^{-2} > \alpha > 5.4 \times 10^{-4}$.
These values are in the same range as those required to match the 
fragments observed with HST and Subaru for 73P.
\citet{desvoivres} present a physical model for the non-gravitational force in terms of ice sublimation and derive from it a range of fragment sizes (times the density, $\rho$ in g~cm$^{-3}$), of $1.4 < \rho a < 24$ m.
Again, these values are comparable to what we derived from the dynamics
and the photometry of the 73P-B fragments in the HST images.

Future high angular resolution imaging of close-approaching comets could reveal how common is ice fragment production and how important a mass loss mechanism is in general.
It is evident that production of large icy fragments is not
exclusively a property of long or short-period comets, given their detection
from C/Hyakutake and 73P/Schwassmann-Wachmann 3.
It is clear that
production of icy fragments is a significant mass loss mechanism, when it is happening. 
For Hyakutake, if we sum only the mass
of the 7 detected fragments, and assume these are the only ones produced
over a month of observation, then the mass production rate is 
$1.2\times 10^5$ g~s$^{-1}$. Of course, it is unlikely that Hyakutake is fragmenting
at the same rate throughout every orbit; if the 7 fragments from 1996 were the only fragments produced
over an entire orbit ($10^5$ yr), then the production rate is a mere 0.1 g~s$^{-1}$.

For 73P-B, from the HST or Subaru images, we estimate at least 20 fragments
with sizes of order 10 m produced over a period less than one week, 
yielding $1.4\times 10^5$ g~s$^{-1}$ mass production rate. 
During the Giotto flyby of comet 26P/Grigg-Skjellerup in 1992, a fragment 
of order 10-100 m was detected \citep{mcbride}, meaning three very different types of
comets have intermediate-sized fragments that are 
too large to have
been levitated from the surface by gas drag, and, 
even if they survive long enough, too small and too far from the nucleus to 
ever be found again as separate comets.
It is not known over what fraction of the orbit such fragmentation occurs, but it
does not seem to be rare.

\subsection{Debris trail and swarm\label{sectrail}}
Debris trails have been detected for more than 30 comets using infrared
observations by {\it IRAS} \citep{sykeswalker} and {\it Spitzer} \citep{reachtrail}.
A debris trail due to 73P is clearly evident in {\it Spitzer} images.
The debris trail from this comet was evident in {\it IRAS} \citep{sykeswalker} 
and {\it COBE} \citep{lisse98} data
before the 1995 splitting event.
Only infrared images have revealed the debris trail; it has not yet been detected
optically despite extensive observations with a wide range of telescopes in 2006.
Debris trails are due to particles in the mm to cm size range, and the 
orbit-averaged mass loss rate is of order $10^4$ g~s$^{-1}$, comparable to the
mass production rate of H$_2$O. 
Averaged over entire orbits, the mass loss rate in the form of debris appears
to be much larger than the mass loss rate in the form of fragments.

The properties of the diffuse infrared emission in the {\it Spitzer} images indicate that the
particles are somewhat different in origin than typical debris trails.
While the surface brightness along the orbit and {\it ahead} of fragment C,
in the 2006 Apr, 2006 May, and 2007 Jan images, is typical of debris trails, the brightness
{\it following} fragment C, and in particular the
significant enhancement of diffuse emission behind fragment B in 2007 Jan,
is highly unusual and cannot be explained by the same origin as typical cometary debris.
For a quiescent comet, the production rate increases toward perihelion, modulo active
region seasonal effects such as are important for 2P/Encke \citep{reachEncke}. The
mm and cm-sized debris gradually separate from the comet; they remain close to the nucleus
during the comet's perihelion passage (due to the small ejection velocity and small
effect of radiation pressure). These large particles form debris trails, or equivalently, meteoriod
streams \citep{jenniskens}, that
follow the comet's orbit on the {\it subsequent} revolutions. 
In this model, the brightness profile along the trail is relatively smooth, gradually decreasing behind
the nucleus and occupying more and more of the comet's orbit until a gravitational perturbation
(normally by Jupiter) changes the comet's orbit and forces the comet to start anew in building its 
meteoroid stream.

\begin{figure}
\plotone{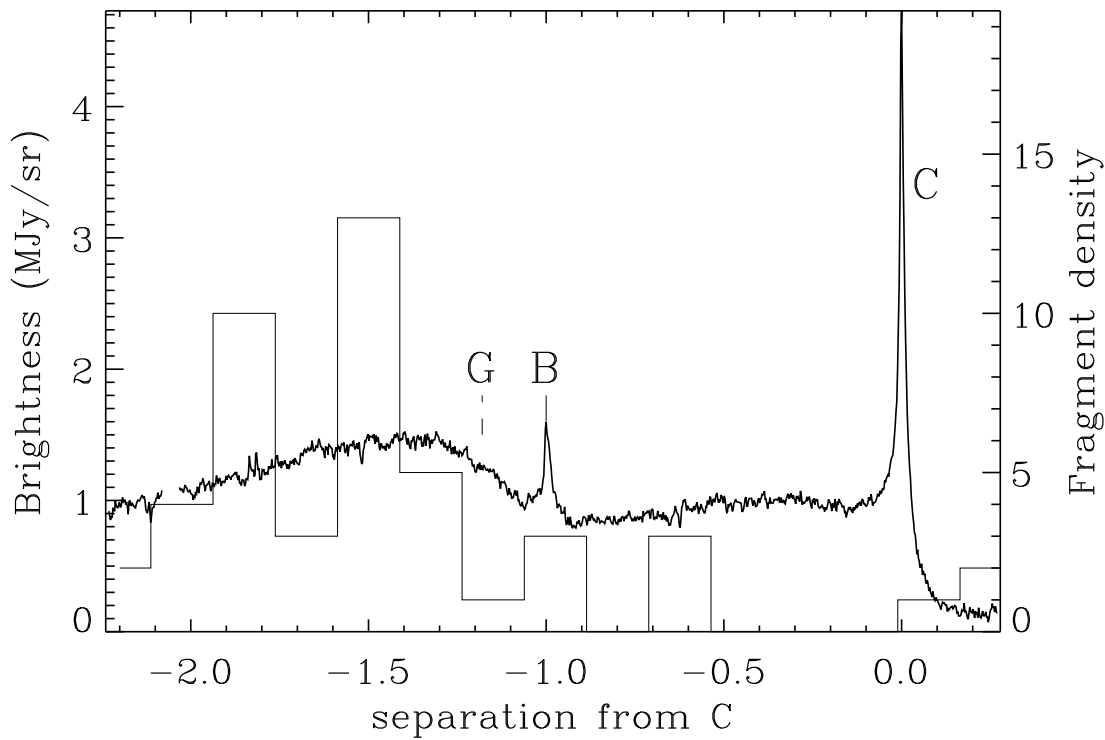}
\figcaption{Surface brightness profile, at wavelength 24 $\mu$m, along the orbit of comet 73P in 2007 Jan (solid curve
and left axis).
The bright fragments B and C are labeled. The relative location expected for fragment G is indicated by a dashed line;
lack of a detected condensation indicates fragment G nearly completely disrupted.
The nonzero brightness ahead of fragment C is the forward debris trail. 
For comparison, the histogram (and right axis) shows the density of fragments detected in 2006 May. The 
fragment density is the number of fragments with
orbital separations, from Table~\ref{fragtab}, within bins every 17\% of the separation between fragments B and C.
\label{profjan}}
\end{figure}

For comet 73P, the brightness profile through the 2007 Jan image is neither monotonic nor smooth. 
Instead, as shown in Figure~\ref{profjan}, there is an elevated brightness behind fragment C that remains fairly constant, then
a significant rise behind fragment B followed by a gradual decline. The broad `bump' in the
surface brightness profile behind fragment B is centered around -1.4 times the separation  between 
fragments B and C ($D_{BC}$), 
relative to fragment C; the width of the bump is approximately 0.3 (FWHM) in the same units.
Could this `bump' be due to debris produced by the swarm of fragments that was detected
in the 2006 May close approach? The distribution of infrared-detected fragments from Table~\ref{fragtab} is
shown in Figure~\ref{profjan} directly on top of the 2007 Jan surface brightness profile. 
It is evident that the distribution of fragments (from 2006 May), scaled to the relative separation between
fragments B and C, is very similar to the debris surface brightness profile; in particular, the high density of fragments, 
approximately 1.5 times $D_{BC}$ behind fragment C, is very likely associated with the surface brightness `bump' in 2007 Jan.
This patch of enhanced brightness is likely a debris field that is 
much of what remains of the fragments that disintegrated during the 2006 perihelion passage.

It is unlikely that diffuse emission
in the 2007 Jan image is due to debris that was produced in the 1995 splitting event.
This is evident from the sheer amount of material required to produce the observed debris field.
In the 2006 Apr image, the surface brightness along the orbit, between fragments B and C, is approximately 0.2 MJy~sr$^{-1}$,
which corresponds to an optical depth of $8.1\times 10^{-10}$ if we take into account a particle temperature
of $278/r_h$ K, where $r_h$ is the distance from the Sun. In the 2007 Jan image, the brightness in the same location
relative to fragments B and C was 0.9 MJy~sr$^{-1}$, which corresponds to an optical depth of $1.1\times 10^{-8}$. Thus
there is 14 times more material between fragments B and C in 2007 Jan than there was in 2006 Apr. (The situation is
likely to be even more extreme in the region behind fragment B, but this region was not known at the time of designing 2006 Apr
observations so that portion of the orbit was not observed and cannot be directly compared.)

The mass of debris in the 2007 Jan image is $3\times 10^{13} \rho a_{\rm cm}$ g, where $\rho$ is the mass
density (g~cm$^{-3}$) of individual particles and $a_{\rm mm}$ is the particle radius in mm. 
The particle size is likely of order mm size, or else radiation pressure would perturb their orbits significantly;
however, the 2007 Jan viewing geometry does not completely rule out a contribution from smaller particles.
The observed width of the debris field perpendicular to the orbit ($4.7\times 10^4$ km FWHM), 
and particle age of 8 months, indicate an expansion velocity of 1.1 m~s$^{-1}$. If the expansion is only due to
the momentum imparted to the particles in the gas outflowing from a nucleus, $v_{ej}\sim 2.3 (\rho a_{\rm cm})^{-1/2}$ m~s$^{-1}$
\citep{sykeswalker,lisse98,reachEncke},
then particle sizes $\rho a_{\rm cm}\sim 4$ are inferred for the debris.
If the apparent separation of a particle from the projected orbit is due to radiation pressure, a lower limit to the particle size can
be inferred; we find $\beta<0.01$ (or equivalently, $\rho a_{\rm cm}>0.006$). 
The {\it location} of the `bump' in the trail profile also contains clues for the
origin of the particles and their size. If the particles originated from fragment B in 2006 May, then in order to reach
their present location at -1.15 $D_{BC}$, they must have $\rho a_{\rm cm}\simeq 0.006$. If the particles originate from the
large, disrupted fragment G, at original location -1.18 $D_{BC}$ in 2006 May, and they moved to -1.4 $D_{BC}$ by 
2007 Jan, then they have $\rho a_{\rm cm}\simeq 0.06$. If the particles originated from the smaller fragment cloud already
near -1.4 $D_{BC}$ in 2006 May and moved less than 0.1 $D_{BC}$ by 2007 Jan, then we require $\rho a_{\rm cm}>0.1$.

The observed debris mass of $\sim 10^{13}$ g produced by fragment disintegration during the 2006 perihelion passage is
equivalent to the entire mass of a body 280 m in diameter. Fragment B has an estimated 400 m diameter, and fragment G was
undoubtedly smaller. Thus the mass of the mm to cm-sized debris was comparable to the entire mass of fragment G
before its disintegration.
The debris mass is also
comparable to or larger than masses of entire meteoroid streams for other comets \citep[cf.][]{reachtrail}.
However, there are clear distinctions between the 73P debris and most other comets.
First, the 73P material in the 2007 Jan image was primarily produced during the 2006 perihelion passage, 
whereas meteoroid streams for comets with stable orbits are built over multiple revolutions.
Second, the 73P material originates from small fragments, whereas for most comets the material 
originates from the nucleus.

\def\extra{
Using the extreme case of 73P, where the extensive field of fragments was well observed due to the close
approach to the Earth, we can consider whether some other comets' debris fields are due to fragment
disintegration rather than levitation of large particles by outflowing material from the primary nucleus.
Of the comets surveyed in the infrared, comet 71P/Clark is the most likely case of a debris field from
fragments. Figure 5 of the {\it Spitzer} comet survey \citep{reachtrail} shows that the peak in the
surface brightness is located far behind the nucleus. Based on this we suggest that 71P was accompanied by
small fragments (which have never been detected individually) that disintegrated in its 2006 perihelion passage.
Besides debris field imaging observations, enhanced infrared brightness of the a comet may be another
indicator of fragment disintegration. The {\it COBE} observations of 73P in 1990 showed it was $\sim 2$ magnitudes brighter
than expected for the nucleus alone, which could mean the comet was already
accompanied by fragments or the comet was already hyperactive preceding its major splitting in 1995 \citep{lisse98}.
}

\subsection{Non-gravitational forces}
Inspecting the available calculations of non-gravitational parameters for
other periodic comets, we found a trend suggesting a correlation
such that comets with high non-gravitational forces 
in their present orbit were more likely to have
split in the past. We found this result using the small body 
database\footnote{http://ssd.jpl.nasa.gov} 
of the JPL Solar System Dynamics group, together with the historical
notes by 
Kronk\footnote{http://cometography.com}
and Kinoshita\footnote{http://www9.ocn.ne.jp/$\sim$comet}. 
For all orbits with measured
non-gravitational forces, we combined the parameters into one
number, $A=10^9 (A_{1}^{2}+A_{2}^{2}+A_{3}^{2})^{1/2}$. Of 43 comets with
$A>3.6$, eight had historically observed splits;
while of the 50 comets with $A< 3.6$, only 2 had
historically recorded splits. 
For the orbits compiled by P. Rocher\footnote{http://www.imcce.fr/page.php?nav=en/ephemerides/donnees/comets/index.php}, 
the logarithmic mean $A=17$
for 7 split comets, while $A=4.7$ for 65 other comets.
Some of the orbits with the highest 
non-gravitational parameters are uncertain, so this result deserves
further scrutiny. But the general tendency is unlikely to change. 

The elevated non-gravitational forces for fragments of cometary splitting
could be due to their smaller size or the presence of freshly exposed interior ice,
both of which are expected properties of fragments. Alternatively, comets with
large non-gravitational forces may be on the verge of splitting due to greater
stresses on the nucleus from the elevated outgassing. For 73P, there are
orbits with measured non-gravitational terms in 1979, 1995, and 2001; the
values $10^9 (A_1,A_2)$ are $(6.7,0.4)$, $(6.9,0.5)$, and $(12.8,1.8)$, 
respectively. Compared to the ensemble of comets, the non-gravitational forces were 
moderate in 1979 and similar during the revolution when the splitting occurred (in 1995).
Then in the first post-splitting revolution, the non-gravitational forces were
much higher. These observations support both hypotheses for the correlation
between non-gravitational forces and splitting, so it is possible that both 
mechanisms (enhanced outgassing leading to splitting, as well as enhanced
outgassing from fragments) may be operating.

\section{Conclusions}

During 2006 March to 2007 Jan,  we used the IRAC and MIPS instruments on the {\it Spitzer} Space Telescope to study the infrared emission from the ensemble of fragments, meteoroids, and dust from
comet 73P/Schwassmann-Wachmann 3 within a
debris field more than 3 degree wide.
We also investigated contemporaneous ground based and HST observations. 
The {\it Spitzer} data spatially resolve the 
fragments' comae and tails on scales of $10^{3}$ km.
The luminosity distribution of large fragments is similar
to the size distribution found for nuclei of the ensemble of Jupiter family comets,
while there is a paucity of smaller fragments.
From fragments 73P-B and C, we use the IRAC 4.5 $\mu$m band 
to measure the 
fluorescence from outflowing CO$_2$ gas, finding that 
the abundance of CO$_{2}$ is 5--10 \% of that of H$_{2}$O.

\begin{deluxetable}{llrcllll}
\tablecaption{Summary of fragment and debris size and dynamical influences\label{sumtable}}
\tabletypesize{\footnotesize}
\tablehead{
\colhead{} & \colhead{} & \multicolumn{2}{c}{Physical properties} &
\colhead{} &\multicolumn{3}{c}{Dynamical parameters} \\
\cline{3-4} \cline{6-8}
\colhead{} & \colhead{location} & \colhead{size} & \colhead{$f_{ice} Z_{16}$} & \colhead{~} & \colhead{$\beta\equiv F_{rad}/F_{grav}$} & 
\colhead{$\alpha\equiv F_{rocket}/F_{grav}$} & 
\colhead{$\nu\equiv v_{ej}/v_{orb}$}
}
\startdata
Largest fragments & along orbit & 200 m     & 0.1   && $10^{-9}$ & $10^{-7}$  & $10^{-6}$ \\ 
Fragments         & along orbit &  50 m     & 0.1   && $10^{-8}$ & $10^{-6}$  & $10^{-6}$ \\
Mini-fragments    & dust tail   &   5 m     & 8     && $10^{-7}$ & $10^{-3}$  & $10^{-6}$ \\
meteoroids        & along orbit &  5 mm  &$<10^{-4}$&& $10^{-4}$ & $<10^{-4}$ & $10^{-4}$ \\
dust              & dust tail   & 10 $\mu$m & $?$   && $10^{-1}$ & $?$        & $10^{-2}$ 
\enddata
\end{deluxetable}

We explain observed morphologies in terms of non-gravitational forces,
which we quantify using the ratio of solar radiation pressure over
gravity, $\beta$, the ratio of momentum loss due to
sublimation over gravity, $\alpha$, and the ratio of 
initial ejection energy to gravity, $\nu$.
Table~\ref{sumtable} summarizes the different types of debris and their
dynamical influences.
The large fragments are splayed along the project orbit of
the progenitor comet with dynamics dominated by $\nu$. 
While the distribution of fragments perpendicular to the orbit is
tightly peaked to the progenitor's orbit, some fragments are
far from the orbit requiring very high $A_{3}$ from non-gravitational
forces or a extremely large ejection velocity perpendicular to the orbit. 
Very small fragments or very large meteoroids, in the $\sim 5$ m 
size range, are seen in ground-based and HST images splayed along
the dust tail; their dynamics are dominated by rocket forces with
$\alpha> 10^{-3}$. 
Smaller meteoroids, in the mm to cm size range, are seen in
the {\it Spitzer} images only; their dynamics are dominated
by weak radiation pressure, $\beta$. If the trail particles were
icy, they would have large $\alpha$ and would deviate significantly
from the debris trail. Thus the debris trail particles must be 
devolatilized.

\clearpage

\acknowledgements 

WTR thanks Giovanni Fazio (Harvard), who as IRAC Principal Investigator supported
part of this project through his Guaranteed Time allotment.
We thank Jeonghee Rho (Caltech) for assistance in generating the
MIPS 24 $\mu$m mosaics. This work is based on observations made with the Spitzer Space
Telescope, which is operated by the Jet Propulsion Laboratory, California
Institute of Technology under a contract with NASA. Support for this work was
provided by NASA through an award issued by JPL/Caltech.
JV thanks the supercomputer facility staff for their support (CINES, France and San Diego Supercomputer Center [SDSC], USA).

\bibliography{wtrbib}

\end{document}